\documentclass[english,11pt,aps,prd,a4paper,preprintnumbers,floatfix,nofootinbib,showpacs,superscriptaddress, notitlepage]{revtex4-1} 
\usepackage[mode=buildnew]{standalone}
\usepackage[usenames,dvipsnames]{color} 
\usepackage{graphicx}
\usepackage{bm}
\usepackage{dcolumn}
\usepackage[colorlinks=true,citecolor=darkred,urlcolor=darkred, pdfborder={0 0 0}]{hyperref}
\usepackage{setspace}
\usepackage{caption}
\usepackage{subcaption}
\usepackage{slashed}
\usepackage[normalem]{ulem}
\usepackage{float}
\usepackage[usenames,dvipsnames]{xcolor}
\usepackage{microtype}
\usepackage{lipsum}
\usepackage{amsmath}
\usepackage{amssymb}
\usepackage{mathrsfs}
\usepackage{placeins}
\usepackage{lettrine}
\usepackage{tikz}
\usetikzlibrary{shapes.misc}
\input Eileen.fd

\definecolor{darkred}{rgb}{0.6,0,0}

\definecolor{linkcolor}{rgb}{0,0,0.5}
\usepackage[T1]{fontenc} 
\usepackage[compat=1.1.0]{tikz-feynhand}
\usepackage{tikz-feynman}
\tikzfeynmanset{compat=1.1.0}
\usepackage{feynmp}
\usepackage{tikzsymbols}
\usepackage{array}
\usepackage{pifont}

\definecolor{linkcolor}{rgb}{0,0,0.5}
\usepackage{orcidlink}
\usepackage{multirow}

\begin{document}
\title{\boldmath \color{Blue}
Generalized chiral $U(1)_{B-L}$: Toward a less constrained $Z'$ and dark matter
}

\author{Ranjeet Kumar\orcidlink{0000-0002-7144-7606}}
\email{kumarranjeet.drk@gmail.com}
\affiliation{Institute for Convergence of Basic Studies, Seoul National University of Science and Technology, Seoul 01811, Republic of Korea}
\author{Hemant Kumar Prajapati\orcidlink{0000-0001-5104-9427}}\email{hemant19@iiserb.ac.in}
\affiliation{Department of Physics, Indian Institute of Science Education and Research - Bhopal \\ Bhopal Bypass Road, Bhauri, Bhopal 462066, India}
\begin{abstract}
  \vspace{1cm} 
  \noindent
Extensions of the Standard Model by a  $U(1)_{X}$ gauge symmetry predict a new $Z'$ boson, whose mass is severely constrained by high-mass dilepton resonance searches at the LHC. We show that these bounds can be relaxed by generalizing the chiral $(B-L)$ charge assignment while preserving gauge anomaly cancellation.  
The resulting $U(1)_X$ charges are parameterized by two independent parameters,
and an appropriate choice of these parameters suppresses the $Z'$ branching fraction into charged leptons, substantially weakening the dilepton constraints. 
The framework accommodates Dirac neutrino masses through a Dirac type-I seesaw mechanism and a stable singlet scalar DM candidate without requiring an additional discrete symmetry.
We find that the modified charge assignment, although beneficial for collider phenomenology, introduces a tension between obtaining the observed relic abundance and satisfying direct detection limits when DM annihilation proceeds solely through the $Z'$ portal. 
The inclusion of scalar-mediated annihilation channels resolves this tension and opens up a broad viable parameter space, with DM masses ranging from $M_{\rm DM}\simeq 60~{\rm GeV}$ to $10~{\rm TeV}$. The resulting parameter space is consistent with the observed relic density, direct detection, collider, and electroweak precision constraints, demonstrating that a generalized chiral charge structure can simultaneously accommodate a less constrained $Z'$ sector, neutrino masses, and viable dark matter phenomenology.

\end{abstract}
\maketitle
%

\section{Introduction}
\label{sec:intro}

Despite its remarkable success, the Standard Model (SM) cannot account for neutrino masses or the nature of dark matter (DM). The discovery of neutrino oscillations established that neutrinos are massive~\cite{Kamiokande-II:1990wrs,Kamiokande-II:1992hns,Super-Kamiokande:1998kpq,Cleveland:1998nv,SNO:2002tuh}, while the origin of their small masses remains an open question. Furthermore, the existence of DM is supported by a wide range of astrophysical and cosmological observations~\cite{Zwicky:1933gu,Rubin:1970zza,Rubin:1980zd,Planck:2018vyg}. 
Together, the evidence for neutrino masses and DM provides a hint toward physics beyond the SM (BSM).
 This has motivated the exploration of a wide range of extensions that can simultaneously address both puzzles.
 In this context, theories with an additional $U(1)_X$ gauge symmetry offer an interesting framework with potentially rich phenomenological implications~\cite{Davidson:1978pm,Mohapatra:1980qe,Mohapatra:1980de,Masiero:1982fi,Mohapatra:1982xz,Buchmuller:1991ce,Leike:1998wr,Appelquist:2002mw,Carena:2004xs,Khalil:2006yi,Montero:2007cd,Basso:2008iv,Oda:2015gna,Rodejohann:2015lca,Das:2016zue,Accomando:2016sge,Bonilla:2017lsq,Alonso:2017uky,Jana:2019mez,Allanach:2018vjg, DeRomeri:2023ytt,Mandal:2023oyh,Ghosh:2024cxi,Majumdar:2024dms,Prajapati:2024wuu,Prajapati:2026tfv,Batra:2026tzz,Kang:2026lgr}

An extension of the SM by an additional $U(1)_X$ gauge symmetry requires the associated fermion charge assignments to satisfy the gauge anomaly cancellation conditions. 
These conditions depend on the chiral structure of the fermion sector. Vectorlike fermions, for which the left- and right-handed components carry identical $U(1)_X$ charges, do not contribute to the net gauge anomalies. In contrast, chiral fermions can lead to nonvanishing anomaly coefficients.
If the SM fermions themselves are charged under $U(1)_X$, their Yukawa interactions and mass generation mechanism impose further restrictions on the allowed charge assignments. A number of anomaly free $U(1)_X$ symmetries have been explored, including $(B-L)$, ($B-3L_i)$, $(B_i-L_j)$, and $(L_i-L_j)$, as well as their linear combinations~\cite{Davidson:1978pm,Mohapatra:1980qe,Mohapatra:1980de,Ma:1997nq,He:1990pn,Khalil:2008kp,Okada:2010wd,Lee:2010hf,Ma:2014qra,Ma:2015raa,Ma:2015mjd}. 
Among these possibilities, the gauged $U(1)_{B-L}$ is particularly well motivated and provides a simple extension of the SM. Its anomaly cancellation naturally requires the introduction of right-handed neutrinos, thereby offering a direct connection to the origin of neutrino masses.

Within the gauged $U(1)_{B-L}$ framework, different choices of the $(B-L)$ charges of the right-handed neutrinos lead to distinct anomaly-free realizations. In particular, two well-studied possibilities are the vector and chiral $(B-L)$ scenarios.
In the vector case, the three right-handed neutrinos carry universal charges (-1,-1,-1), whereas in the chiral realization their charges are assigned as (-4,-4,5) \cite{Montero:2007cd,Ma:2014qra}.
The chiral case is of particular interest because the enlarged neutrino charges alter the partial decay widths of the associated $Z'$ gauge boson, enhancing its invisible decay width. 
The resulting suppression of the dilepton branching fraction can, in turn, weaken the collider constraints on the $Z'$. 
The strongest such bounds come from searches for heavy
dilepton resonances at the LHC~\cite{ATLAS:2019erb,CMS:2021ctt},
which are quoted in terms of
$\sigma(pp\to Z')\times{\rm BR}(Z'\to\ell^{+}\ell^{-})$, and are therefore sensitive to the dilepton branching fractions as well as to the $Z'$ production rate.
This motivates us to ask whether a further generalization of the chiral charge structure can suppress the dilepton branching fraction, thereby allowing a less constrained $Z'$ sector. 

In this work, we study a generalization of the chiral $(B-L)$ framework. We extend the SM gauge symmetry by an additional $U(1)_X$ and derive the anomaly-free charge assignments in the presence of three right-handed neutrinos. 
Imposing the gauge anomaly cancellation conditions together with gauge invariance of the Yukawa interactions, we find that all fermion charges can be expressed in terms of two free parameters, $X_L$ and $\kappa$. In particular, the right-handed neutrinos carry charges $(-4\kappa,-4\kappa,5\kappa)$.
The conventional chiral $(B-L)$ solution is recovered for $X_L= -1$ and $\kappa=1$. Suitable choices of these parameters can further enhance the invisible width of the $Z'$, thereby suppressing its dilepton branching fraction and weakening the LHC bounds.
Beyond the gauge sector phenomenology, the generalized $U(1)_X$ framework also accommodates neutrino masses and a stable DM candidate.
The $U(1)_X$ symmetry is spontaneously broken by an additional scalar singlet $\chi$, whose VEV generates the $Z'$ mass and plays a crucial role in neutrino mass generation.

In the neutrino sector, the effective operator $\overline{L}\tilde{\Phi}\nu_{R_\alpha}\chi$ ($\alpha=1,2$) generates Dirac neutrino masses, while the right-handed neutrino with charge $5\kappa$ remains decoupled from this mechanism, resulting in one massless neutrino state. 
At the renormalizable level, this operator arises from a Dirac type-I seesaw mediated by two pairs of BSM fermions, $(N_{L_p},N_{R_p})$, with $p=1,2$.
For the DM sector, we introduce a singlet scalar DM candidate $\chi_d$ whose stability follows from the $U(1)_X$ charge assignment, without requiring an additional discrete symmetry.
We investigate both the pure $Z'$-portal limit and the scenario with additional scalar-mediated annihilation channels. While the enhanced fermion charges that suppress the dilepton branching fraction lead to a tension between the relic density and direct detection constraints in the pure gauge limit, the inclusion of the scalar sector provides additional annihilation mechanisms and substantially enlarges the viable DM parameter space. In particular, we find viable DM masses ranging from approximately $60$ GeV to $10$ TeV while satisfying the observed relic density, direct detection, collider, and electroweak precision constraints.

The paper is organized as follows. In Sec.~\ref{sec:anomaly}, we derive the anomaly cancellation conditions and obtain the general fermion charge assignment. The model construction and the realization of neutrino masses through the Dirac type-I seesaw mechanism are presented in Sec.~\ref{sec:model}. In Sec .~\ref{sec:Zprime}, we investigate the $Z'$ phenomenology and the resulting collider constraints. The DM phenomenology of the model is studied in Sec.~\ref{sec:dm}. Finally, our numerical findings and concluding remarks are summarized in Sec.~\ref{sec:conc}.

\section{Gauge Anomaly cancellation conditions} \label{sec:anomaly}

We begin with the discussion of the gauge anomaly cancellation conditions in the presence of the additional $U(1)_X$ gauge symmetry. 
The consistency of any chiral gauge theory requires the cancellation of all gauge and mixed gauge-gravitational anomalies. While the SM fermion content is anomaly-free with respect to the SM gauge group, the extension by an extra $U(1)_X$ symmetry generally introduces new anomaly constraints. Consequently, the $U(1)_X$ charge assignments of both the SM and BSM fermions must satisfy a set of anomaly cancellation conditions to ensure the theory remains unitary and renormalizable~\cite{Adler:1969gk,Bardeen:1969md,Bell:1969ts,Delbourgo:1972xb,Alvarez-Gaume:1983ihn,Witten:1982fp}. 
The corresponding anomaly cancellation conditions can be written as follows \cite{Prajapati:2024wuu,Prajapati:2026tfv},
\begin{subequations}\label{Eq:Anomaly_can_con}
\begin{align}
&[SU(3)_{C}]^2[U(1)_X]= \left(2 X_{Q} -  X_{u_{R}}-X_{d_{R}}\right) \label{Eq:anoUx1} = 0\,, 
\\
&[SU(2)_{L}]^2[U(1)_X]= \left(X_{L} +3 X_{Q}\right) = 0\label{Eq:anoUx2}\,,
\\
&[U(1)_{Y}]^2 [U(1)_X]=  \left( X_{L} + \frac{1}{3}  X_{Q} -2  X_{e_{R}} -\frac{8}{3}X_{u_{R}}-\frac{2}{3}  X_{d_{R}} \right) = 0\label{Eq:anoUx3}\,, 
\\
&[U(1)_{Y}] [U(1)_X]^2=  \left\{   \left(X_{Q}\right)^{2}-\left(X_{L}\right)^{2}  + \left(X_{e_{R}}\right)^2 -2 \left(X_{u_{R}}\right)^2 + \left(X_{d_{R}}\right)^2  \right\} = 0 \label{Eq:anoUx4} \,,
\\ 
& [U(1)_X]^3=  \left[ 2\left(X_{L}\right)^{3}  +6 \left(X_{Q}\right)^{3}   - \left(X_{e_{R}}\right)^{3} -3 \left\{ \left(X_{u_{R}}\right)^{3}  + \left(X_{d_{R}}\right)^{3} \right\} \right]  - \sum_{i=1}^{3} \left(X_{\nu_{R_i}}\right)^{3}/3 = 0 \label{Eq:anoUx5} \,,
\\&[G]^2[U(1)_X]=  \left\{ 2X_{L}  + 6 X_{Q} -X_{e_{R}}  -3\left( X_{u_{R}} + X_{d_{R}}\right) \right\}  - \sum_{i=1}^{3} X_{\nu_{R_i}}/3 = 0 \label{Eq:anoUx6}\,.
\end{align}
\end{subequations}
%
Here $X_{\psi}$ denotes the $U(1)_X$ charge of the fermion $\psi$. The
left-handed quark and lepton doublets are $Q\equiv(u_L,d_L)^{T}$ and
$L\equiv(\nu_L,e_L)^{T}$, while $u_R$, $d_R$ and $e_R$ are the corresponding
$SU(2)_L$ singlets. We also introduce three right-handed neutrinos $\nu_{R_i}$ ($i = 1,2,3$), which are singlets under the SM gauge group. 
For the theory to be anomaly-free, all six conditions in Eq.~\eqref{Eq:Anomaly_can_con}, together with the anomaly cancellation conditions associated with the SM gauge sector, must be satisfied simultaneously. A detailed discussion of the anomaly cancellation conditions in the SM can be found in Ref.~\cite{Prajapati:2024wuu}.


Apart from the gauge anomalies, the chiral nature of the SM fermions under $U(1)_X$
requires care in writing down the Yukawa interactions. 
The anomaly cancellation
conditions must therefore be supplemented by the requirement that the Yukawa couplings remain gauge invariant. 
Imposing this on the SM Yukawa interactions responsible for
fermion mass generation yields additional relations among the $U(1)_X$ charges, reducing the number of independent parameters. The SM Yukawa Lagrangian reads
\begin{equation}\label{Eq:SM_Yukawa_couplings}
   - \mathcal{L}_{Y} = Y_{e}^{ij}\overline{L}_{i} \Phi e_{R_j} +Y_{u}^{ij} \overline{Q}_{i} \tilde{\Phi} u_{R_j}  + Y_{d}^{ij} \overline{Q}_{i} \Phi d_{R_j} + \text{H.c.}\,,  
\end{equation}
where $\tilde{\Phi} = i\sigma_2 \Phi^*$. Requiring Eq.~\eqref{Eq:SM_Yukawa_couplings} to
be invariant under the additional $U(1)_X$ imposes the following constraints on the
$U(1)_X$ charge of the Higgs doublet $\Phi$,
\begin{equation}\label{Eq:Higgs_Charge_Constraints}
    X_{\Phi} = \frac{X_{\psi_{R}} - X_{\psi_{L}}}{2 T^{3}_{\psi_{L}}}\,,
\end{equation}
where $T^{3}_{\psi_{L}}$ represent the third component of the weak isospin. 
Consequently, one obtains the following relations,
\begin{equation} \label{Eq:Higgs_Charge_Final}
X_{\Phi}=X_{L}-X_{e_{R}}=X_{Q}-X_{d_{R}}=X_{u_{R}}-X_{Q}.
\end{equation}
Solving the first four anomaly cancellation conditions of Eq.~\eqref{Eq:Anomaly_can_con} together with the Yukawa invariance relations of Eq.~\eqref{Eq:Higgs_Charge_Final} gives a unique solution for the $U(1)_{X}$ charges of the SM fermions,
\begin{equation}\label{Eq:anomalycancel_4_eq}
X_{Q}=-\frac{X_{L}}{3},~X_{u_{R}}=X_{\Phi} - \frac{X_{L}}{3},~X_{d_{R}}= -\left(X_{\Phi} + \frac{X_{L}}{3} \right),~X_{e_{R}}=X_{L}-X_{\Phi}.
\end{equation}
Interestingly, the $U(1)_X$ charges of all SM fermions can be expressed as a linear combination of the SM hypercharge and $(B-L)$ charges,
\begin{equation}\label{Eq:General_charges}
X_{\psi} = X_{\Phi} Y_{\psi} -(X_{L}+X_{\Phi})(B-L)_{\psi}.
\end{equation}
Here $Y_{\psi}$ and $(B-L)_{\psi}$ is the hypercharge and the $(B-L)$ charge of the fermion $\psi$ defined in Table \ref{tab:particles}. 
Hence, we are left with two free parameters $X_{L}$ and $X_{\Phi}$, along with $U(1)_{X}$ charges of three right-handed fermions $X_{\nu_{R_{i}}}$. The relation between them is given by Eq. \eqref{Eq:anoUx5} and Eq. \eqref{Eq:anoUx6}. Solving these two equations, using Eq. \eqref{Eq:anomalycancel_4_eq} gives,
\begin{subequations}\label{lowcondition}
\begin{align}
\sum_{i=1}^{3} (X_{\nu_{R_i}})^{3}=& ~ 3 (X_{L} + X_{\Phi})^{3},\\
\sum_{i=1}^{3} X_{\nu_{R_i}}=& ~ 3 (X_{L} + X_{\Phi}).
\end{align}
\end{subequations}

The conditions in Eq.~\eqref{lowcondition} do not uniquely determine the $U(1)_X$ charge assignments of the three right-handed neutrinos. Instead, they admit several classes of solutions, each corresponding to a distinct set of $U(1)_X$ charges satisfying both the anomaly cancellation and Yukawa invariance conditions \cite{Prajapati:2024wuu}. 
The nature of these charge assignments for right-handed neutrinos determines the Yukawa interactions and mass terms allowed by the gauge symmetry, thereby dictating the viable mechanism for neutrino mass generation. In the following section, we present one interesting charge assignment and discuss its implications for neutrino mass generation.

\newpage

\section{Model Framework} \label{sec:model}
Having discussed the anomaly-free charge assignments in the previous section, we now discuss the corresponding model framework and its implications for neutrino mass generation.
Among the various anomaly-free solutions satisfying Eq.~\eqref{lowcondition}, we focus on a particularly well-motivated class corresponding to the generalized chiral $U(1)_{B-L}$ scenario. 
The framework is based on the widely studied chiral $(B-L)$ model, where the three right-handed neutrinos carry the non-universal charges $(-4,-4,5)$~\cite{Montero:2007cd,Ma:2014qra,Ma:2015mjd}. Here, we generalize this charge assignment by introducing an arbitrary parameter $\kappa$, such that the corresponding charges are given by $(-4\kappa,-4\kappa,5\kappa)$~\cite{Prajapati:2024wuu}.
Besides providing a viable framework for neutrino mass generation, this generalized solution gives rise to distinctive phenomenological signatures. 
Specifically, the enlarged, non-universal charge assignments of the right-handed neutrinos can substantially enhance the invisible decay width of the $Z'$ gauge boson. As we will demonstrate in the following sections, this feature elegantly suppresses the dileptonic branching fraction, thereby significantly relaxing the stringent heavy resonance constraints imposed by current LHC searches ~\cite{CMS:2018mgb,ATLAS:2019erb,CMS:2021ctt}.
Beyond its implications for the $Z'$ phenomenology, this generalized gauge structure also has important consequences for the DM phenomenology.
In view of these phenomenological implications, we focus on this class of solutions in the following discussion. 
The corresponding $U(1)_X$ charge assignments of the SM particles and the right-handed neutrinos are summarized in Table~\ref{tab:General_Charges}, where these charges are parameterized in terms of two free parameters, namely $\kappa$ and $X_{L}$.
\footnote{These charges can equivalently be obtained from the general expression in Eq.~\eqref{Eq:General_charges}, $X_{\psi} = X_{\Phi}Y_{\psi}-(X_{\Phi}+X_L)(B-L)_{\psi} = X_{\Phi}Y_{\psi}+\kappa\,(B-L)_{\psi}$, by assigning the right-handed neutrinos the $(B-L)$ charges $(-4,-4,5)$.}
The chiral $(B-L)$ solution corresponds to the special case $X_{L}=-1$ and $\kappa=1$, for which $X_{\Phi}=0$ and the right-handed neutrino charges reduce to $(-4,-4,5)$.
\begin{table}[!h]
\centering
\renewcommand{\arraystretch}{1.5}
\setlength{\tabcolsep}{10pt}
\begin{adjustbox}{max width=\textwidth}
\begin{tabular}{|c|c|c|c|c|c|c|c|c|}
\hline
$Q$ &
$u_{R}$ &
$d_{R}$ &
$L$ &
$e_{R}$ &
$\nu_{R_1}$ &
$\nu_{R_2}$ &
$\nu_{R_3}$ &
$\Phi$ \\
\hline\hline
$\displaystyle -\frac{X_{L}}{3}$ &
$\displaystyle -\frac{4X_{L}}{3}-\kappa$ &
$\displaystyle \kappa+\frac{2X_{L}}{3}$ &
$\displaystyle X_{L}$ &
$\displaystyle \kappa+2 X_{L}$ &
$\displaystyle -4\kappa$ &
$\displaystyle -4\kappa$ &
$\displaystyle  5\kappa$ &
$\displaystyle -X_{L}-\kappa$ \\
\hline
\end{tabular}
\end{adjustbox}
\caption{Anomaly-free $U(1)_X$ charge assignments of the SM fermions, the right-handed neutrinos $\nu_{R_i}$, and the Higgs doublet $\Phi$ in the generalized chiral $U(1)_{B-L}$ framework, parametrized by $X_L$ and $\kappa$. The chiral $(B-L)$ model is recovered for $X_L=-1$, $\kappa=1$.}
\label{tab:General_Charges}
\end{table}

The generalized $U(1)_X$ framework can be extended to accommodate neutrino masses by introducing additional particles without altering the anomaly-free charge structure.
In our model framework, neutrino masses are generated through the effective operator $\overline{L}\tilde{\Phi}\nu_{R_\alpha}\chi$ ($\alpha=1,2$) whose UV completion is realized via a Dirac type-I seesaw mechanism. Here, $\chi$ is an SM singlet scalar with $U(1)_X$ charge $3\kappa$. The UV completion of this operator requires two vector-like fermion pairs, $(N_{L_p},N_{R_p})$, with $U(1)_X$ charge $-\kappa$, where $p=1,2$. Thus, only the right-handed neutrinos $\nu_{R_{1,2}}$, carrying the charge $-4\kappa$, participate in neutrino mass generation, while $\nu_{R_3}$, with charge $5\kappa$, remains decoupled from the neutrino sector.\footnote{The third right-handed neutrino $\nu_{R_3}$ can also be incorporated into the neutrino mass generation mechanism by extending the scalar sector with appropriately charged fields, allowing the corresponding effective operators to satisfy the $U(1)_X$ gauge symmetry.} In addition, we introduce an SM singlet scalar $\chi_d$ as the DM candidate, carrying an arbitrary $U(1)_X$ charge $q_{\rm DM}$. The particle content of the model and the corresponding charge assignments under the different symmetries are summarized in Table~\ref{tab:particles}.
The $U(1)_X$ charge of $\chi_d$ can be chosen such that all interactions that could induce its decay are forbidden by the gauge symmetry. Consequently, the $U(1)_X$ symmetry naturally ensures the stability of the DM candidate without the need to impose an \textit{ad hoc} discrete symmetry.
In this work, we choose $q_{\rm DM}=5/2$, with $X_L=-1$ and $\kappa=2$, as discussed in the subsequent section. This choice leads to the $U(1)_X$ charges of $\chi$ and $\Phi$ being $6$ and $-1$, respectively, which forbid the possible decay interactions of the DM candidate and ensure its stability. However, this property does not hold for an arbitrary choice of $q_{\rm DM}$. The charge assignment must be chosen consistently with the $U(1)_X$ charges of the other fields, particularly $\chi$ and $\Phi$, to forbid all allowed DM decay interactions.

\begin{center}
\begin{table}[!t]
\centering
\renewcommand{\arraystretch}{1.6}
\setlength{\tabcolsep}{12pt}
\begin{tabular}{|c|c|c|c|}
\hline
 & Fields & $SU(3)_C \otimes SU(2)_L \otimes U(1)_Y$ & $U(1)_{X}$ \\
\hline \hline
\multirow{6}{*}{\rotatebox{90}{SM}}
 & $Q$      & $(3, 2, \frac{1}{3})$  & $-\dfrac{X_L}{3}$ \\
 & $u_R$    & $(3, 1, \frac{4}{3})$  & $-\dfrac{4X_{L}}{3}-\kappa$ \\
 & $d_R$    & $(3, 1, -\frac{2}{3})$ & $\kappa+\dfrac{2X_{L}}{3}$ \\
 & $L$      & $(1, 2, -1)$           & $X_{L}$ \\
 & $e_R$    & $(1, 1, -2)$           & $\kappa+2 X_{L}$ \\
 & $\Phi$   & $(1, 2, 1)$            & $-X_{L}-\kappa$ \\
\hline
\multirow{4}{*}{\rotatebox{90}{BSM}}
 & $\nu_{R_i}$           & $(1, 1, 0)$ & $(-4\kappa,\,-4\kappa,\,5\kappa)$ \\
 & $N_{L_p},\,N_{R_p}$   & $(1, 1, 0)$ & $-\kappa$ \\
 & $\chi$                & $(1, 1, 0)$ & $3\kappa$ \\
 & $\boldsymbol{\chi_{d}}$ & $\mathbf{(1, 1, 0)}$ & $\boldsymbol{q_{\rm DM}}$ \\
\hline
\end{tabular}
\caption{Particle content of the model and the corresponding transformation properties under the SM gauge group and $U(1)_X$. The index $p=1,2$ labels the two vector-like fermion pairs.}
\label{tab:particles}
\end{table}
\end{center} 

\subsection{Yukawa sector and neutrino mass generation}

\begin{figure}[!h]
    \centering
    \input {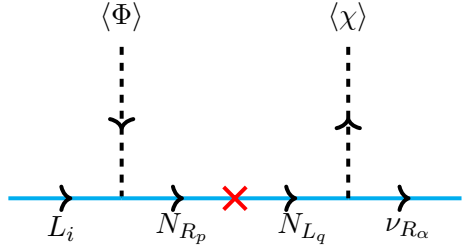}
    \caption{Dirac neutrino mass generation via the type-I seesaw mechanism, with $i=1,2,3$ and $p,q,\alpha=1,2$.}
    \label{fig:feyn_numass}
\end{figure}

In this section, we briefly discuss the Yukawa sector and the mechanism responsible for generating neutrino masses. The relevant interactions are mediated by the vector-like fermion pair $(N_L,N_R)$ and, after integrating out these heavy states, give rise to the effective operator $\overline{L}\tilde{\Phi}\nu_{R_\alpha}\chi$. This operator is realized through a Dirac type-I seesaw mechanism, as illustrated by the Feynman diagram in Fig.~\ref{fig:feyn_numass}.  
Once the scalars $\Phi$ and $\chi$ acquire their respective VEVs, the relevant symmetry breaking generates Dirac neutrino masses. The corresponding Yukawa Lagrangian is given by 
\begin{equation}
    -\mathcal{L}^{\nu}_Y = Y_{\nu}^{i p}\, \overline{L}_i\Tilde{\Phi}N_{R_p} + M^{q p} \, \overline{N}_{L_q}N_{R_p} + Y_{\chi}^{q \alpha} \, \, \overline{N}_{L_q}\nu_{R_\alpha} \, \chi + \text{H.c.}\,.
\end{equation}
The neutrino mass matrix, in the basis $(\overline{\nu}_{L_i},\overline{N}_{L_q} )$ and $({\nu_{R_\alpha}},{N_{R_p}} )^T$, takes the form
\begin{align}
    \mathcal{M}_{\nu} = \begin{pmatrix}
         0 & \mathcal{M_{D}} \\
         \mathcal{M'_{D}} & \mathcal{M}
     \end{pmatrix},
\end{align}
where, $\mathcal{M_{D}} = Y^{ip}_{\nu}v_{\Phi}/\sqrt{2} $, $\mathcal{M'_{D}} = Y^{q \alpha}_{\chi}v_{\chi}/\sqrt{2} $, and $\mathcal{M} = M^{qp}$. 
Taking the seesaw limit, $M^{qp} \gg Y^{ip}_{\nu}v_{\Phi}, Y^{q\alpha}_{\chi}v_{\chi} $, the resulting light Dirac neutrino mass matrix is given by
\begin{align}
    m_{\nu} = -\mathcal{M_{D}} \mathcal{M}^{-1} \mathcal{M'_{D}} = -\frac{v_{\Phi} v_{\chi}}{2} Y^{ip}_{\nu} (M^{qp})^{-1} Y^{q\alpha}_{\chi} \ .
\end{align}
Here, $Y^{ip}_{\nu}$ is $3\times 2$ matrix, $M^{qp}$ is $2\times 2$ matrix, and $Y^{q \alpha}_{\chi}$ is $2\times 2$ matrix. Thus, the resulting neutrino mass matrix $m_{\nu}$ is $3\times 2$ matrix with rank 2. Consequently, the model predicts two massive light neutrinos and one massless state, a minimal neutrino mass spectrum that remains compatible with current neutrino oscillation data \cite{Kamiokande-II:1990wrs,Kamiokande-II:1992hns,Super-Kamiokande:1998kpq,Cleveland:1998nv,SNO:2002tuh}.
We would like to mention that the masslessness of the third neutrino is a consequence of the minimal scalar content adopted in our construction rather than a fundamental requirement of the anomaly-free gauge structure.
In particular, an additional scalar singlet with $U(1)_X$ charge $-6\kappa$ could be introduced to generate a mass for the third neutrino through the corresponding effective operator. We do not pursue this extended scalar sector in the present work.


\subsection{Scalar sector and mass spectrum}\label{subsec:Scalar}
We now turn to the scalar sector, which consists of the SM Higgs doublet $\Phi$ and two SM singlet scalars, $\chi$ and $\chi_d$. The most general renormalizable scalar potential consistent with the symmetries of the model is given by
\begin{align}\label{eq:scalar_Potential}
         \mathcal{V}_{\mathtt{S}} &=  m_{\Phi}^2 \Phi^{\dagger}\Phi + \frac{\lambda_\Phi}{2} (\Phi^{\dagger}\Phi)^{2} + m_{\chi}^2 (\chi^{*} \chi) + \frac{\lambda_{\chi}}{2} (\chi^{*} \chi)^2  + m_{\chi_{d}}^2 (\chi_{d}^{*} \chi_{d}) + \frac{\lambda_{\chi_{d}}}{2} (\chi_{d}^{*} \chi_{d})^2 + \lambda_{\Phi \chi} (\Phi^{\dagger}\Phi) (\chi^{*} \chi) \nonumber \\& ~~~~+ \lambda_{\Phi \chi_{d}} (\Phi^{\dagger}\Phi)(\chi_{d}^{*} \chi_{d}) + \lambda_{\chi \chi_{d}}(\chi^{*} \chi)(\chi_{d}^{*} \chi_{d})\,.
\end{align}
The requirement that the scalar potential be bounded from below at tree level leads to the following conditions on the quartic couplings~\cite{Kannike:2012pe}:
\begin{align}
&\lambda_{\Phi} > 0, \quad 
\lambda_{\chi} > 0, \quad 
\lambda_{\chi_d} > 0,
\\[2mm]
&
\lambda_{\Phi\chi}+\sqrt{\lambda_{\Phi}\lambda_{\chi}} > 0, \
\lambda_{\Phi\chi_d}+\sqrt{\lambda_{\Phi}\lambda_{\chi_d}} > 0, \
\lambda_{\chi\chi_d}+\sqrt{\lambda_{\chi}\lambda_{\chi_d}} > 0, 
\\[2mm]
&\sqrt{\lambda_{\Phi}\lambda_{\chi}\lambda_{\chi_d}}
+\lambda_{\Phi\chi}\sqrt{\lambda_{\chi_d}}
+\lambda_{\Phi\chi_d}\sqrt{\lambda_{\chi}}
+\lambda_{\chi\chi_d}\sqrt{\lambda_{\Phi}}
\nonumber\\
&\quad
+\sqrt{2
\left(\lambda_{\Phi\chi}+\sqrt{\lambda_{\Phi}\lambda_{\chi}}\right)
\left(\lambda_{\Phi\chi_d}+\sqrt{\lambda_{\Phi}\lambda_{\chi_d}}\right)
\left(\lambda_{\chi\chi_d}+\sqrt{\lambda_{\chi}\lambda_{\chi_d}}\right)}
>0 .
\end{align}
We further require the quartic couplings to remain within the perturbative regime, which we impose through the condition given by
\begin{align}
\left|\lambda_{\Phi}\right| &< \sqrt{4\pi}, \quad
\left|\lambda_{\chi}\right| < \sqrt{4\pi}, \quad
\left|\lambda_{\chi_d}\right| < \sqrt{4\pi},
\nonumber\\
\left|\lambda_{\Phi\chi}\right| &< \sqrt{4\pi}, \quad
\left|\lambda_{\Phi\chi_d}\right| < \sqrt{4\pi}, \quad
\left|\lambda_{\chi\chi_d}\right| < \sqrt{4\pi} .
\end{align}

We next discuss the scalar mass spectrum subject to the theoretical constraints on the quartic couplings. Upon spontaneous symmetry breaking, the scalar fields can be written as
\begin{equation}\label{eq:Field_Expansion}
    \Phi = \begin{pmatrix}
        G^{+} \\ \frac{v_{\Phi} + h + iG^{0}}{\sqrt{2}} 
    \end{pmatrix}, \quad     
    \chi = \frac{1}{\sqrt{2}} (v_{\chi} + R_{\chi} + i I_{\chi})\,.
\end{equation}
By solving the minimization conditions, the mass parameters $m_{\Phi}^2$ and $m_{\chi}^2$ can be expressed as
\begin{equation}\label{eq:Tadpole}
\begin{split}
    & 2 m_{\Phi}^{2} + \lambda_{\Phi}v_{\Phi}^2 +  \lambda_{\Phi \chi} v_{\chi}^{2} = 0, \\
    & 2 m_{\chi}^{2}  + \lambda_{\chi} v_{\chi}^{2} + \lambda_{\Phi \chi} v_{\Phi}^2 = 0 .
\end{split}    
\end{equation}
The scalar $\chi_d$ does not acquire a VEV and remains stable, thereby providing a viable DM candidate within our model framework.
The fields $G^{\pm}$, $G^{0}$ and $I_{\chi}$ are the massless Goldstone bosons absorbed by the $W^{\pm}$, $Z$, and $Z'$ respectively. This leaves $h$ and $R_{\chi}$ as the physical CP-even scalars. In the $(h,R_{\chi})^{T}$ basis their mass matrix reads
\begin{equation}\label{Eq:App:Higgs_Mixing}
    \mathcal{M}^{2}_{H} = \begin{pmatrix}
        \lambda_{\Phi}v_{\Phi}^2 & v_{\Phi} v_{\chi} \lambda_{\Phi \chi} \\
        v_{\Phi} v_{\chi} \lambda_{\Phi \chi} & \lambda_{\chi}v_{\chi}^2
    \end{pmatrix}\, \equiv 
    \begin{pmatrix}
        A & C  \\
        C & B
    \end{pmatrix}\,.
\end{equation}
The mass eigenvalues of light and heavy mass eigenstates are given by
\begin{align}
M^{2}_{H_{1}} & =\frac{1}{2}\left[A+B-\sqrt{(A-B)^2+4C^2}\right], \\
M^{2}_{H_{2}} & =\frac{1}{2}\left[A+B+\sqrt{(A-B)^2+4C^2}\right],
\end{align}
both eigenvalues remain positive provided
$\lambda_{\Phi}\lambda_{\chi}>\lambda_{\Phi\chi}^{2}$.
We follow the convention $M_{H_{1}} < M_{H_{2}}$ and have identified $H_{1}$ as the SM Higgs, with mass $M_{H_{1}}=125$~GeV. The two mass eigenstates $H_{1}, H_{2}$ are related with the $(h, R_{\chi})$ fields through the following rotation matrix 
\begin{equation}
\begin{bmatrix}
H_{1} \\
H_{2}  \end{bmatrix} = \begin{bmatrix}
\cos\theta_{h} & -\sin\theta_{h} \\
\sin\theta_{h} & \cos\theta_{h}
\end{bmatrix} \begin{bmatrix}
h \\
R_{\chi}
\end{bmatrix}, \,\, \text{with}\,\, \tan 2\theta_{h}=\frac{2C}{B-A}.
\end{equation} 
The mass of the DM candidate $\chi_d$ is determined by its bare mass parameter $m_{\chi_d}$ and the VEVs of $\Phi$ and $\chi$, and can be expressed as
\begin{equation}
    M_{\rm DM}^{2}= m_{\chi_{d}}^{2} + \frac{1}{2}( \lambda_{\Phi \chi_{d}} v_{\Phi}^{2} + \lambda_{\chi \chi_{d} }v_{\chi}^{2})\,.
\end{equation}
The resulting scalar spectrum has important implications for the phenomenology discussed in the subsequent sections.

\subsection{Gauge sector}
We next examine the gauge sector of the model, which is extended by an additional neutral gauge boson $Z'$ associated with the gauged $U(1)_X$ symmetry. If the Higgs doublet carries a non-zero $U(1)_X$ charge, the $Z'$ mixes with the neutral SM gauge bosons following the spontaneous breaking of the electroweak and $U(1)_X$ symmetries. We therefore analyze the resulting neutral gauge boson mass spectrum.
Upon spontaneous symmetry breaking, the gauge boson masses are generated through the scalar kinetic terms. The relevant kinetic terms can be written as
\begin{equation}\label{Eq:Scalar_Kinetic}
(D_{\mu}\Phi)^{\dagger}D^{\mu} \Phi + (D_{\mu}\chi)^{\dagger}D^{\mu}\chi \,,
\end{equation}
where the covariant derivative is given by 
\begin{equation}\label{EQ:COD}
D_{\mu}= \partial_{\mu} + igT^{a}_{w}W^{a}_{\mu} + ig' \frac{Y}{2} B_{\mu}+ig_{x}XC_{\mu}\,.
\end{equation}
The gauge couplings $g$, $g'$ and $g_{x}$ correspond to $SU(2)_{L}$,
$U(1)_{Y}$ and $U(1)_{X}$, with $Y$ and $X$ the hypercharge and $U(1)_{X}$
charge of the field on which $D_{\mu}$ acts. The $SU(2)_{L}$ generators are
$T^{a}_{w}=\sigma^{a}/2$, where $\sigma^{a}$ are the Pauli matrices.
Once the electroweak and $U(1)_{X}$ symmetries are spontaneously broken, the
scalar fields acquire the VEV,
\begin{equation}\label{EQ:VEVS}
\langle \Phi \rangle = \frac{1}{\sqrt{2}}\begin{pmatrix} 0 \\ v_{\Phi} \end{pmatrix},
\qquad
\langle \chi \rangle = \frac{v_{\chi}}{\sqrt{2}}\,.
\end{equation}
Substituting Eqs.~\eqref{EQ:COD} and~\eqref{EQ:VEVS} into the scalar kinetic
terms of Eq.~\eqref{Eq:Scalar_Kinetic} yields the gauge boson mass terms.
For the electrically charged gauge bosons, it is convenient to introduce the
mass eigenstates
$W^{\pm}_{\mu}=\left(W^{1}_{\mu}\mp i\,W^{2}_{\mu}\right)/\sqrt{2}$, with
the corresponding generators
$T_{w}^{\pm}=\left(T_{w}^{1}\pm i\,T_{w}^{2}\right)/\sqrt{2}$.
Since $\chi$ is an $SU(2)_{L}$ singlet, no mixing arises in the charged sector, and the $W^{\pm}$ mass remains identical to its SM expression, $M_{W} = g\, v_{\Phi}/2$.
The neutral gauge bosons, in contrast, mix with one another after spontaneous symmetry breaking. 
In the basis $\left(B_{\mu},\,W^{3}_{\mu},\,C_{\mu}\right)$ their mass matrix takes the following form
\begin{equation}\label{Eq:Gauge:B:Mmat}
\mathcal{M}^2_{_{V}}= \frac{v_{\Phi}^{2}}{4}\begin{pmatrix}
g'^{2} & -gg' & 2g'X_{_{\Phi}}\,g_{x}\\
-gg'   &  g^{2} & -2gX_{_{\Phi}}\,g_{x}\\
2g'X_{_{\Phi}}\,g_{x} & -2gX_{_{\Phi}}\,g_{x} & 4u^{2}\,g_{x}^{2}
\end{pmatrix}\,,
\end{equation}
where $u^{2}=X^{^{2}}_{_{\Phi}}+9 \kappa^{2} (v_{\chi}/v_{\Phi})^{2}$, and $X_{_{\Phi}}=-X_{L} - \kappa$ denotes the $U(1)_X$ charge of the SM Higgs doublet. 
The mass matrix in Eq. \eqref{Eq:Gauge:B:Mmat} can be diagonalized by an orthogonal matrix $\mathcal{O}_{(\alpha)}$, and the diagonal mass matrix are given as,  $(\mathcal{M}^2_{_{V}})_{\rm dia} = \mathcal{O}_{(\alpha)} \mathcal{M}^2_{_{V}} \mathcal{O}^{^{\dagger}}_{(\alpha)}$.  The mass ($A^\mu, Z^\mu, Z^{\prime \mu}$) and gauge states are related to each other as,
\begin{equation}
\label{Eq:unitary_matrix}
\begin{bmatrix}
A^{\mu} \\
Z^{\mu} \\
Z^{\prime\mu}
\end{bmatrix} =
\begin{bmatrix}
\cos\theta_{W} &~ \sin\theta_{W} &~0\\
-\cos\alpha \sin\theta_{W} & \cos\alpha \cos\theta_{W}
&~ \sin \alpha\\
\sin\alpha \sin\theta_{W} &~  -\sin\alpha\cos\theta_{W} &~ \cos\alpha 
\end{bmatrix}  \begin{bmatrix}
B^{\mu} \\
W_{3}^{\mu}\\
C^{\mu}
\end{bmatrix}.
\end{equation}
Following the diagonalization, one mass eigenstate is zero and is identified with the photon $A^\mu$. 
The other two mass eigenstates are given by,
\begin{equation}\label{mass}
M_{Z}^{2}= \frac{v_{\Phi}^{2}}{8}(A_{0}-\sqrt{B_{0}^{2}+C_{0}^{2}}),\,M_{Z'}^{2}=\frac{v_{\Phi}^{2}}{8}(A_{0}+\sqrt{B_{0}^{2}+C_{0}^{2}})\,,
\end{equation}
where, $A_{0}=g^{2}+{g'}^{2}+4u^{2}g_{x}^{2},~ B_{0}=4X_{_{\Phi}}g_{x}\sqrt{g^{2}+{g'}^{2}}$, $C_{0}=4u^{2}g_{x}^{2}-(g^{2}+{g'}^{2})$\,. 
The VEV of the singlet scalar $\chi$ can be expressed in terms of the $Z'$
mass as
\begin{equation}\label{Eq:ux_interm_Mzp}
    v_{\chi} = \frac{M_{Z'}}{3 \kappa \, g_x} \sqrt{\frac{4M_{Z'}^{2} -v_{\Phi}^{2}(g^{2} +g'^{2} + 4X_{\Phi}^{2}\,g_{x}^{2}) }{4M_{Z'}^{2} - v_{\Phi}^{2}(g^{2}+g'^{2})}} \,.
\end{equation}
The two mixing angles appearing in Eq.~\eqref{Eq:unitary_matrix} are given by\footnote{Here we assume the Higgs $U(1)_X$ charge to be positive, so that $\alpha$ remains positive. If the Higgs charge is negative, one can equivalently replace $\alpha \rightarrow -\alpha$ in the rotation matrix, i.e.,
$\mathcal{O}_{(\alpha)} \rightarrow \mathcal{O}_{(-\alpha)}$,
thereby keeping $\alpha$ positive.},
\begin{equation}\label{Eq:Rotation_angle}
    \tan\theta_{W} = \frac{g'}{g},\quad \sin 2\alpha = \frac{4X_{\Phi}g_x}{ \sqrt{g^{2}+{g'}^{2}} } \frac{(M_{Z}^{\text{SM}})^{2}}{ M_{Z'}^{2}-M_{Z}^{2}} \,,
\end{equation}
where $\theta_{W}$ is the usual SM weak mixing angle and $M_{Z}^{\rm SM}$
denotes the tree-level SM $Z$ boson mass,
$M_{Z}^{\rm SM}=v_{\Phi}\sqrt{g^{2}+g'^{2}}/2$.
The angle $\alpha$ depends on the $U(1)_{X}$ charge of the Higgs doublet $X_{\Phi}$, the gauge coupling $g_{x}$ and the $Z'$ mass $M_{Z'}$.
Since the mixing is induced by the Higgs doublet, it vanishes in the limit $X_{\Phi}\to0$.
This is the case for the chiral $(B-L)$ model, where
$X_{L}=-1, \ \kappa=1$ gives $X_{\Phi}=0$ and hence no $Z-Z'$ mixing. 
In the limit $\alpha\to0$ the correction to the $Z$ boson mass also vanishes and the $Z$ boson mass
reduces to its SM value, $M_{Z}\to M_{Z}^{\rm SM}$. For $\alpha\neq0$ the $Z$ boson mass is always pushed below $M_{Z}^{\rm SM}$, while $M_{W}$ is
unaffected.
The ratio of the two masses is constrained by the precisely measured
electroweak parameter
$\rho=M_{W}^{2}/\left(M_{Z}^{2}\cos^{2}\theta_{W}\right)$
\cite{Ross:1975fq,Bento:2023weq}, which equals unity at tree level in the
SM. In ${\rm SM}\otimes U(1)_{X}$ theories it instead deviates from unity,
\begin{equation}
\label{Eq:Rho_parameter}
    \rho-1 = \left[\left(\frac{M_{Z'}}{M_{Z}}\right)^{2}-1\right]
    \sin^{2}\alpha\,.
\end{equation}
 \begin{figure}[!h]
     \centering
     \includegraphics[width=0.42\linewidth]{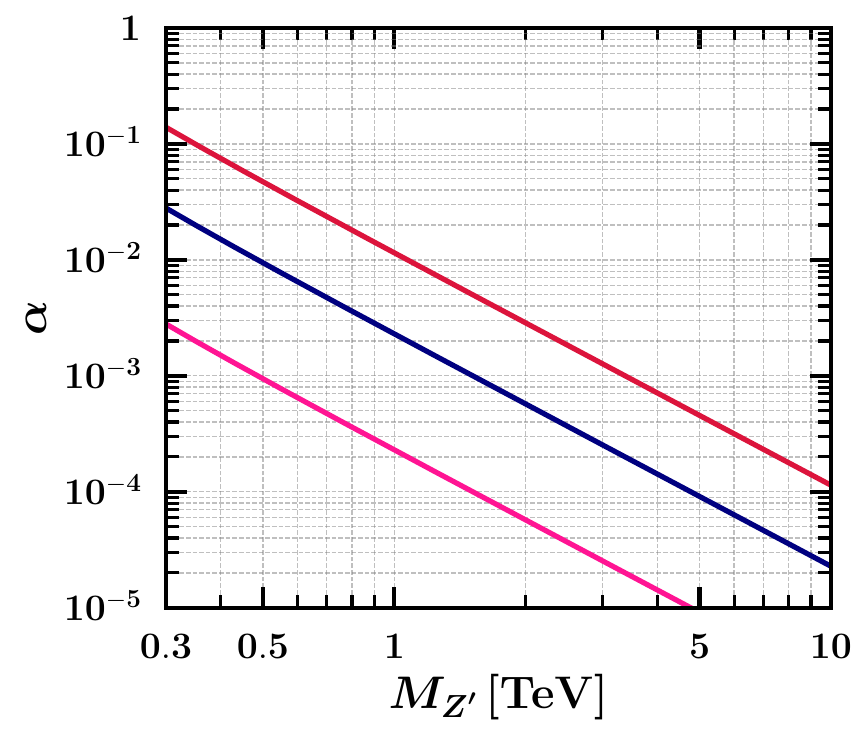}
      \includegraphics[width=0.565\linewidth]{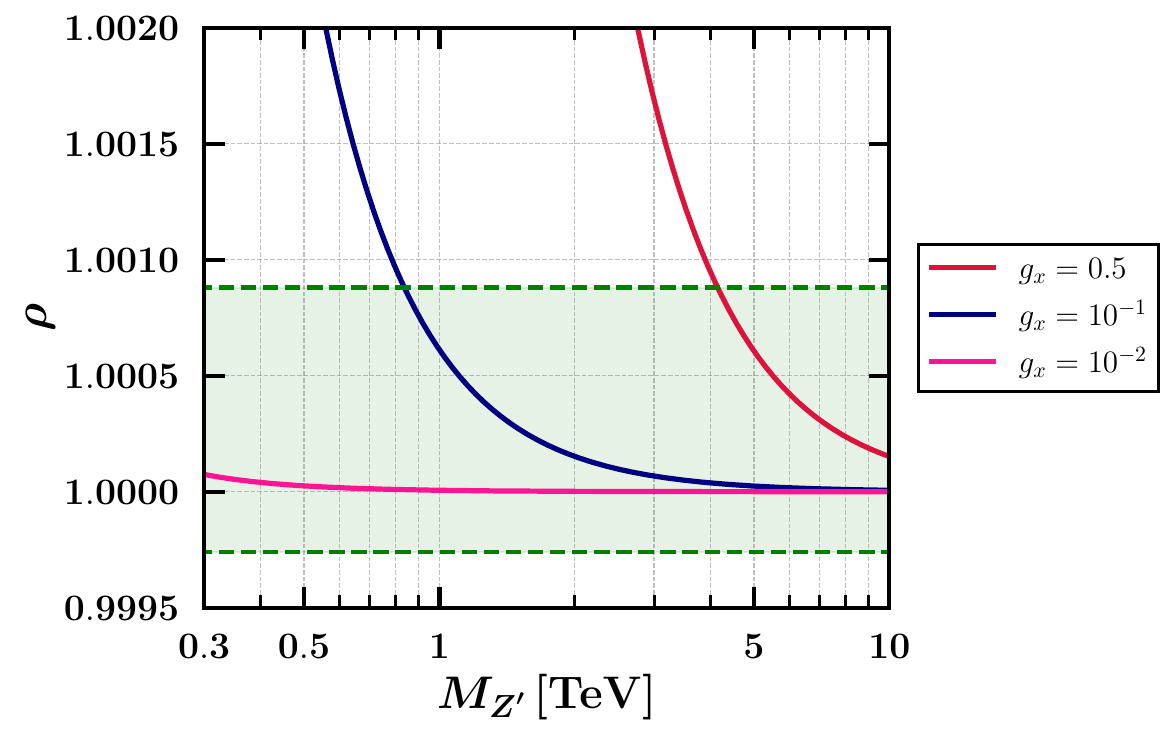}
    \caption{Left: the $Z-Z'$ mixing angle $\alpha$ as a function of $M_{Z'}$
for $g_{x}=0.5$ (red), $10^{-1}$ (blue) and $10^{-2}$ (magenta). Right: the
corresponding tree level $\rho$ parameter obtained from
Eq.~\eqref{Eq:Rho_parameter}, with the green band showing the experimentally
allowed range~\cite{ParticleDataGroup:2024cfk}. For both panels, the Higgs charge is fixed to the  benchmark value $X_{\Phi}=1$.}
     \label{fig:alpha_Rho}
 \end{figure}

The mixing angle $\alpha$ and the resulting $\rho$ parameter are shown as
functions of the $Z'$ mass in Fig.~\ref{fig:alpha_Rho} for three representative
values of the gauge coupling. In both panels, the Higgs charge is fixed to
its benchmark value $X_{\Phi}=1$. The green band indicates the $3\sigma$
range allowed by the most recent global fit to electroweak precision data,
$\rho=1.00031 \pm 0.00057$~\cite{ParticleDataGroup:2024cfk}.
Since the $Z$ boson mass lies below its SM value in this scenario, the
$\rho$ parameter exceeds unity. Large values of $g_{x}$ at low $Z'$ masses are therefore excluded by the $\rho$
parameter, as shown in Fig.~\ref{fig:alpha_Rho}. 
Besides shifting the $Z$ mass, the $Z-Z'$ mixing also modifies the couplings of the $Z$ boson to the SM fermions. We discuss these effects and the resulting constraints from the LEP measurements in the next subsection.
\subsection{Gauge boson couplings to fermions}
We now derive the couplings of the gauge bosons to the SM fermions. Once the neutral gauge fields are rotated to their mass eigenstates (see Eq.~\eqref{Eq:unitary_matrix}), the covariant derivative takes the form
\begin{equation}\label{Eq:codrivative2}
\begin{split}
D_{\mu}= &~ \partial_{\mu} +D_{\mu}^{\text{WCC}} + D_{\mu}^{\text{QED}}+ D_{\mu}^{\text{WNC}} +  D_{\mu}^{\text{BSM}}\,,
\end{split}
\end{equation}
where,
\begin{subequations}
\label{Eq:codrivative3}
\begin{align}
D_{\mu}^{\text{WCC}} &~= ig \left[ T^{+} W_{\mu}^{+} + T^{-}W_{\mu}^{-} \right],\,~~~~ D_{\mu}^{\text{QED}} = ieQA_{\mu}\,,  \\
D_{\mu}^{\text{WNC}} &~= i\frac{g}{\cos \theta_{W}} \left[ (T^{3} - Q\sin^{2} \theta_{W}) \cos \alpha + X\frac{\cos \theta_{W}}{g} g_{x}\sin \alpha   \right] Z_{\mu}\,, \\
 D_{\mu}^{\text{BSM}} &~= i \left[-  \frac{g}{\cos \theta_{W}} (T^{3} - Q \sin^{2} \theta_{W}) \sin \alpha + X g_{x} \cos \alpha \right] Z'_{\mu}\,.
\end{align}
\end{subequations}
Here $T^3$ and $Q$ represent the third component of the weak isospin and the electric charge of a fermion, respectively and $X$ represents its  $U(1)_X$ charge. 
From Eq.~\eqref{Eq:codrivative3}, one can infer that the charged-current interactions mediated by the $W^\pm$ bosons are identical to those of the SM, since no additional mixing arises in the charged sector.
Likewise, the choice of the rotation matrix in Eq.~\eqref{Eq:unitary_matrix} leaves the electromagnetic interactions invariant. 
The neutral-current interactions mediated by the $Z$ and $Z'$ bosons, on the other hand, are modified and take the following form,
\begin{equation} \label{NC_lagrangian}
- \mathcal{L}_{\text{NC}} =  \frac{g}{\cos\theta_{W}}\, \overline{\psi} \gamma^{\mu}\left(g_{\psi_{_{\mathtt{L}}}}^{z} P_{\mathtt{L}} + g_{\psi_{_{\mathtt{R}}}}^{z} P_{\mathtt{R}}\right)\psi\, Z_{\mu}+ \overline{\psi} \gamma^{\mu}\left(g_{\psi_{_{\mathtt{L}}}}^{z'} P_{\mathtt{L}}+g_{\psi_{_{\mathtt{R}}}}^{z'} P_{\mathtt{R}}\right)\psi\, Z'_{\mu} \,. 
\end{equation} 
The couplings $g_{\psi_{_{\mathtt{L/R}}}}^{z}$ and $g_{\psi_{_{\mathtt{L/R}}}}^{z'}$ are given by
\begin{subequations}\label{NC_Couplings}
\begin{align}
&g_{\psi_{_{\mathtt{L}}}}^{z} = \Big(T_{\psi_{_{\mathtt{L}}}}^3 - Q_{\psi}\sin^{2}\theta_{W}\Big)\cos{\alpha} + \frac{X_{\psi_{_{\mathtt{L}}}}g_x}{g}\sin{\alpha}\cos{\theta_{W}}\,,\label{NC_Couplings1} \\
&g_{\psi_{_{\mathtt{R}}}}^{z} = - Q_{\psi}\sin^{2}\theta_{W}\cos{\alpha} + \frac{X_{\psi_{_{\mathtt{R}}}}g_x}{g}\sin{\alpha}\cos{\theta_{W}}\, ,\label{NC_Couplings2}\\
& g_{\psi_{_{\mathtt{L}}}}^{z'} = -\frac{g}{\cos \theta_{W}} \Big(T_{\psi_{_{\mathtt{L}}}}^3 - Q_{\psi} \sin^{2} \theta_{W} \Big)\sin{\alpha} + X_{\psi_{_{\mathtt{L}}}}g_x\cos{\alpha}\,,\label{NC_Couplings3} \\
& g_{\psi_{_{\mathtt{R}}}}^{z'} =   \frac{g}{\cos \theta_{W}}\, Q_{\psi} \sin^{2} \theta_{W}\,\sin{\alpha} + X_{\psi_{_{\mathtt{R}}}}g_x\cos{\alpha}\,.\label{NC_Couplings4}
\end{align}
\end{subequations}
Here, $T_{\psi_{_{\mathtt{L}}}}^3$ and $Q_{\psi}$ represent the third component of the weak isospin and the electric charge of the fermion $\psi$, respectively. 
Note that, owing to the $Z-Z'$ mixing, the couplings of the $Z$ boson to fermions receive a BSM contribution in addition to the SM one. This extra piece is controlled by the gauge coupling $g_x$ and the mixing angle $\alpha$, and vanishes in the limit $\alpha \to 0$. These parameters can therefore be constrained by the effective vector and axial-vector couplings of the neutral weak current at the $Z$ pole, extracted from the electroweak precision observables (EWPOs) measured by the LEP Electroweak Working Group and SLD~\cite{ALEPH:2005ab,ParticleDataGroup:2026mpi,Huang:2019obt}.
To compare our results with these measurements, we take into account the radiative corrections arising from the gauge boson self-energies and the flavor-dependent $Z\bar\psi\psi$ vertex corrections. Including these effects, the vector and axial-vector couplings of the $Z$ boson to a fermion $\psi$ become~\cite{ALEPH:2005ab,ParticleDataGroup:2026mpi}
\begin{subequations}\label{SM_Couplings_Z}
\begin{align}
&g_{V}^{\psi} = \sqrt{\rho_{\psi}}\Big(T_{\psi_{_{\mathtt{L}}}}^3 - 2Q_{\psi}\sin^{2}\theta_{\rm eff}^{\psi}\Big)\cos{\alpha} + \frac{(X_{\psi_{_{\mathtt{L}}}} + X_{\psi_{_{\mathtt{R}}}})\,g_x}{g}\sin{\alpha}\cos{\theta_{W}}\,,\label{SM Couplings1Z} \\
&g_{A}^{\psi} = \sqrt{\rho_{\psi}}\;T_{\psi_{_{\mathtt{L}}}}^3\cos{\alpha} + \frac{(X_{\psi_{_{\mathtt{L}}}} - X_{\psi_{_{\mathtt{R}}}})\,g_x}{g}\sin{\alpha}\cos{\theta_{W}}\,. \label{SM Couplings2Z}
\end{align}
\end{subequations}
Here, $\rho_\psi$ and $\sin^2\theta_{\rm eff}^{\psi} = \kappa_\psi \sin^2\theta_W$ account for the SM radiative corrections; their explicit forms are given in Refs.~\cite{ALEPH:2005ab}.
The leptonic effective couplings were measured at LEP with a precision of $\mathcal{O}(10^{-4})$, with substantially higher precision than the quark couplings, and therefore provide the strongest constraints~\cite{ALEPH:2005ab}.
Deriving these constraints requires fixing the $U(1)_X$ charges of the leptons and of the Higgs doublet. We adopt the benchmark $X_L=-1$, $\kappa=2$, for which $X_{e_R}=0$ and $X_\Phi=-1$. Since $|X_\Phi|=1$, as in Fig.~\ref{fig:alpha_Rho}, the resulting bounds can be directly compared with those from the $\rho$ parameter.
\begin{figure}[ht]
\begin{center}
\includegraphics[width=0.48\linewidth]{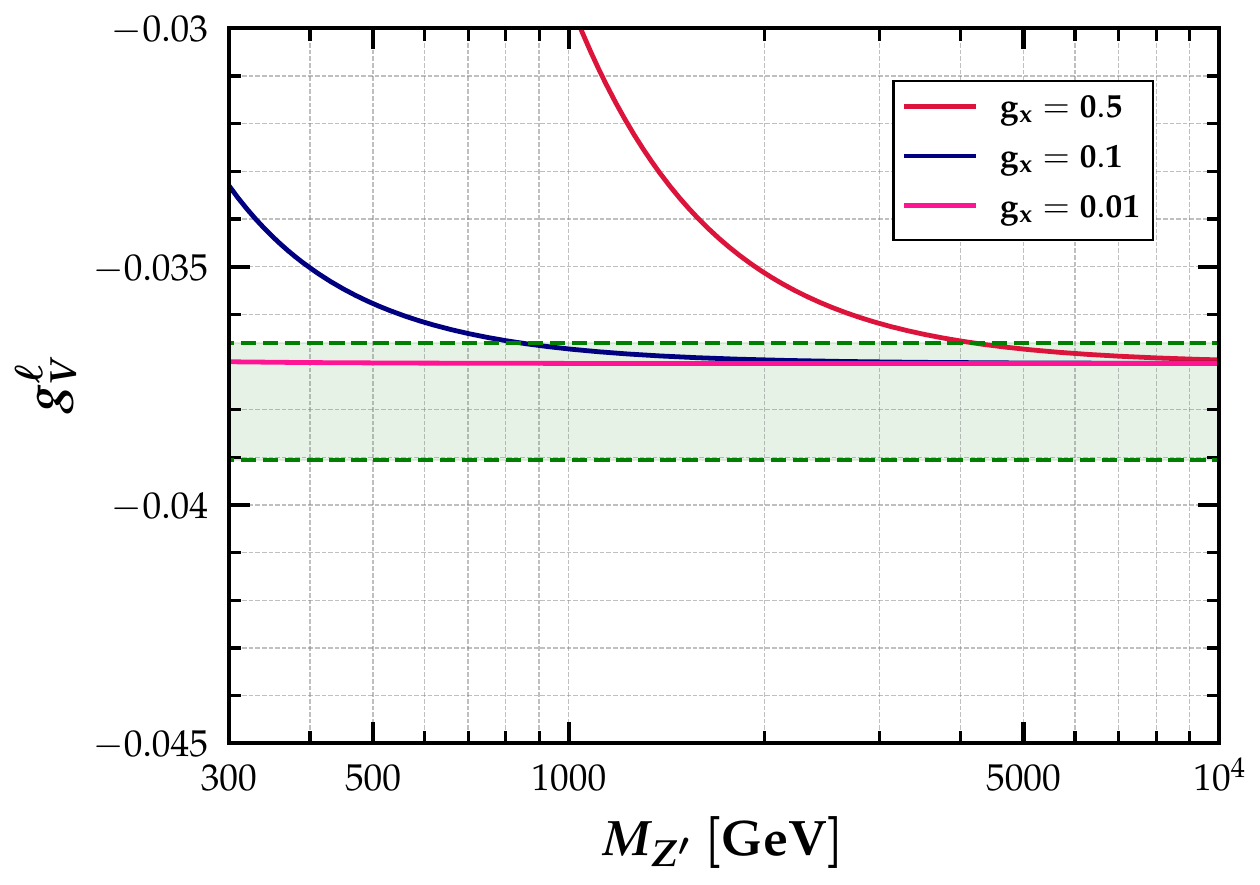}
\includegraphics[width=0.48\linewidth]{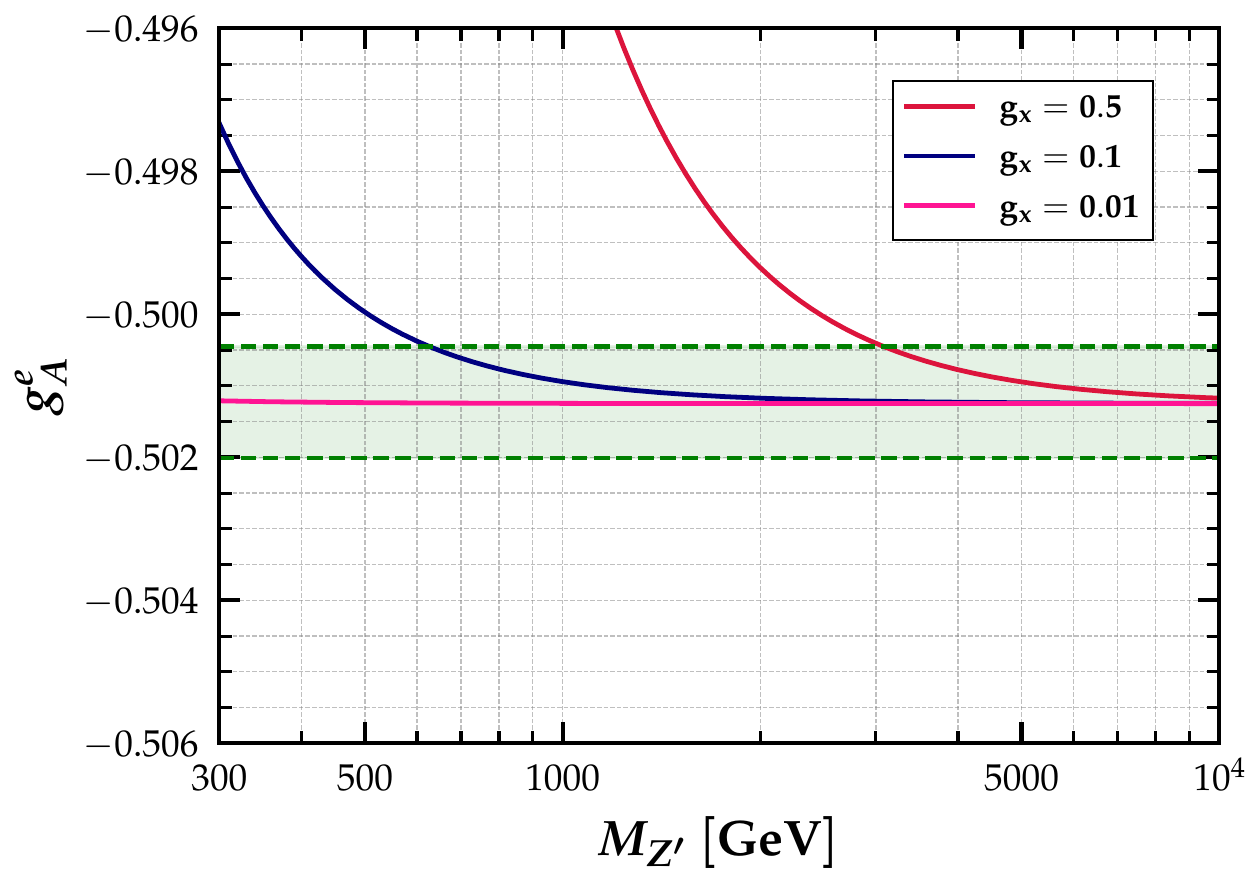}
\end{center}
\caption{Effective vector (left) and axial-vector (right) couplings of the $Z$ boson to charged leptons as functions of $M_{Z'}$, obtained from Eq.~\eqref{SM_Couplings_Z} for the benchmark $X_L=-1$, $\kappa=2$. The red, blue, and magenta curves correspond to $g_x = 0.5$, $0.1$, and $0.01$, respectively. The green shaded bands bounded by dashed lines indicate the $3\sigma$ ranges allowed by the LEP/SLD measurements~\cite{ALEPH:2005ab,ParticleDataGroup:2024cfk}. The predictions approach the SM values as $M_{Z'}$ increases and the $Z-Z'$ mixing decreases.}
\label{FiG:Zp_EFF_Coup}
\end{figure}
In Fig.~\ref{FiG:Zp_EFF_Coup}, we compare the predicted effective vector (left) and axial-vector (right) couplings of the $Z$ boson to charged leptons, given in Eq.~\eqref{SM_Couplings_Z}, with the LEP measurements as functions of $M_{Z'}$, for this benchmark and three representative values of $g_x$.
In evaluating the SM part of the couplings, we fix the effective leptonic mixing angle and the $\rho_\ell$ parameter to their LEP central values, $\sin^2\theta_{\rm eff}^{\ell} = 0.23153$ and $\rho_\ell = 1.0050$~\cite{ALEPH:2005ab}.
As evident from Fig.~\ref{FiG:Zp_EFF_Coup}, the vector coupling $g_V^\ell$ provides the stronger constraint. Since $X_{e_R}=0$ in our benchmark, the BSM shifts of $g_V^\ell$ and $g_A^\ell$ are equal in magnitude; the stronger bound from $g_V^\ell$ arises because the SM prediction, evaluated with the LEP central values of $\sin^2\theta_{\rm eff}^\ell$ and $\rho_\ell$, already lies close to the upper edge of the measured range, and the $Z-Z'$ mixing shifts it further away. These constraints are comparable to those from the $\rho$ parameter discussed in Fig.~\ref{fig:alpha_Rho}.

We emphasize that, apart from the $Z$ pole analysis above, the discussion so far has been kept general with respect to the $U(1)_X$ charge assignments. 
This general framework suffices to construct the gauge and scalar sectors and, in particular, to realize neutrino mass generation. 
Quantitative predictions for the collider and DM phenomenology, however, require a definite choice of the free parameters $X_L$ and $\kappa$. 
We therefore adopt the same benchmark as in the $Z$ pole analysis, $X_L=-1$, $\kappa=2$, and compare it throughout with the chiral $(B-L)$ case, $X_L=-1$, $\kappa=1$, as well as with the vector $(B-L)$ scenario. The resulting charges are listed in Table~\ref{Tab:U(1)_BM_Charges}.
The charge assignment adopted here, $X_{L}=-1$, $\kappa=2$, differs from the chiral $(B-L)$ case, $X_{L}=-1$, $\kappa=1$, only through the value of $\kappa$.
Increasing $\kappa$ from one to two doubles the right-handed neutrino charges, from $(-4,-4,5)$ to $(-8,-8,10)$. At the same time, it renders $e_{R}$ neutral under $U(1)_X$ and assigns a non-zero charge, $X_{\Phi}=-1$, to the Higgs doublet. The right-handed quarks also acquire larger $U(1)_X$ charges than in the chiral and vector $(B-L)$ cases.
These features are responsible for the phenomenological differences discussed in the following sections.
We stress that this choice is made solely for the phenomenological analysis; the general features of the gauge, scalar, and neutrino sectors discussed above are unaffected.
With the charge assignment fixed, we now turn to the phenomenology of the model. In Sec.~\ref{sec:Zprime}, we derive the collider constraints that determine the viable parameter space. We then study the DM phenomenology in Sec.~\ref{sec:dm}, taking the stable scalar $\chi_d$ as the DM candidate with $U(1)_X$ charge $q_{\rm DM}=5/2$.

\section{Collider Phenomenology of the $Z'$} \label{sec:Zprime}
We now investigate the collider phenomenology of the $Z'$ boson, considering constraints from both hadron and lepton colliders. We first focus on the LHC, where searches for heavy resonances in the dilepton (dielectron and dimuon) final state provide stringent constraints on the $Z'$ parameter space \cite{CMS:2018mgb,ATLAS:2019erb,CMS:2021ctt}. We then discuss the complementary constraints from LEP-II.
The $Z'$ boson can be resonantly produced at the LHC through the Drell-Yan process $q\bar{q}\to Z'$, where $q$ denotes either a valence or a sea quark in the proton. Such production requires the quarks to be charged under $U(1)_X$. If the SM Higgs doublet carries a non-zero $U(1)_X$ charge, the resulting $Z-Z'$ mixing provides an additional contribution to the $q\bar{q}Z'$ coupling, suppressed by the mixing angle.\footnote{Kinetic mixing between hypercharge and $U(1)_X$ can also induce $Z-Z'$ mixing in general. We consider kinetic mixing to be small and neglect this contribution for simplicity.} The $Z'$ subsequently decays into any SM or BSM state carrying $U(1)_X$
charge, and the dilepton mode yields the cleanest experimental signature.
These searches provide the most stringent limits on the $Z'$ mass and
coupling up to $M_{Z'}\sim6$~TeV, beyond which the production cross section
falls steeply and the sensitivity is lost.

Both the ATLAS and CMS collaborations have searched for such resonances in the dilepton final state, $q\bar{q}\to Z'\to\ell^{+}\ell^{-}$ \cite{CMS:2018mgb,ATLAS:2019erb,CMS:2021ctt}. No significant deviation from the SM expectation was observed, and the results are presented as bounds on the production cross section times the dilepton branching fraction, $\sigma(pp\to Z')\times{\rm BR}(Z'\to\ell^{+}\ell^{-})$ with $\ell=e,\mu$, as a function of the resonance mass $M_{Z'}$.
Being sensitive to ${\rm BR}(Z'\to\ell^{+}\ell^{-})$, these limits can be
weakened by suppressing the decay of the $Z'$ into charged leptons, which can be
achieved by enlarging its invisible
width~\cite{CMS:2018mgb,ATLAS:2019erb,CMS:2021ctt,Prajapati:2024wuu,Kang:2026lgr}.
In the present model the right-handed neutrinos carry $U(1)_X$ charges and can
contribute substantially to the invisible width of the $Z'$, reducing
${\rm BR}(Z'\to\ell^{+}\ell^{-})$ and thereby relaxing the resulting
constraint on $M_{Z'}$. 
%
To recast these limits in the model, we adopt the benchmark charge
assignment given in the last row of Table~\ref{Tab:U(1)_BM_Charges}, which
follows from the general anomaly-free solution of
Table~\ref{tab:particles} for $\kappa=2$ and $X_{L}=-1$. The charges of the
vector and chiral $(B-L)$ models are also listed for comparison.
\begin{table}[!h]
\centering
\renewcommand{\arraystretch}{2.2}
\setlength{\tabcolsep}{10pt}
\begin{adjustbox}{max width=\textwidth}
\begin{tabular}{|l|c|c|c|c|c|c|c|c|c|}
\hline
\textbf{Models} &
$Q$ &
$u_{R}$ &
$d_{R}$ &
$L$ &
$e_{R}$ &
$\nu_{R_1}$ &
$\nu_{R_2}$ &
$\nu_{R_3}$ &
$\Phi$ \\
\hline\hline
\textbf{Vector} $(B-L)$ &
$\displaystyle \frac{1}{3}$ &
$\displaystyle \frac{1}{3}$ &
$\displaystyle \frac{1}{3}$ &
$-1$ &
$-1$ &
$-1$ &
$-1$ &
$-1$ &
$0$ \\
\hline
\textbf{Chiral $(B-L)$} &
$\displaystyle \frac{1}{3}$ &
$\displaystyle \frac{1}{3}$ &
$\displaystyle \frac{1}{3}$ &
$-1$ &
$-1$ &
$-4$ &
$-4$ &
$5$ &
$0$ \\
\hline
\textbf{This model} &
$\displaystyle \frac{1}{3}$ &
$\displaystyle -\frac{2}{3}$ &
$\displaystyle \frac{4}{3}$ &
$-1$ &
$0$ &
$-8$ &
$-8$ &
$10$ &
$-1$ \\
\hline
\end{tabular}
\end{adjustbox}
\caption{$U(1)_X$ charge assignments of the fermions and of the scalar
$\Phi$ for the three cases considered. The chiral $(B-L)$ and the new model presented follow from the general anomaly-free solution for $X_L=-1, \ \kappa=1$ and $X_L=-1$, $\kappa=2$, respectively, while the vector-like $(B-L)$ case is listed for comparison.}
\label{Tab:U(1)_BM_Charges}
\end{table}

\begin{figure}[!h]
     \centering
     \includegraphics[width=0.6\linewidth]{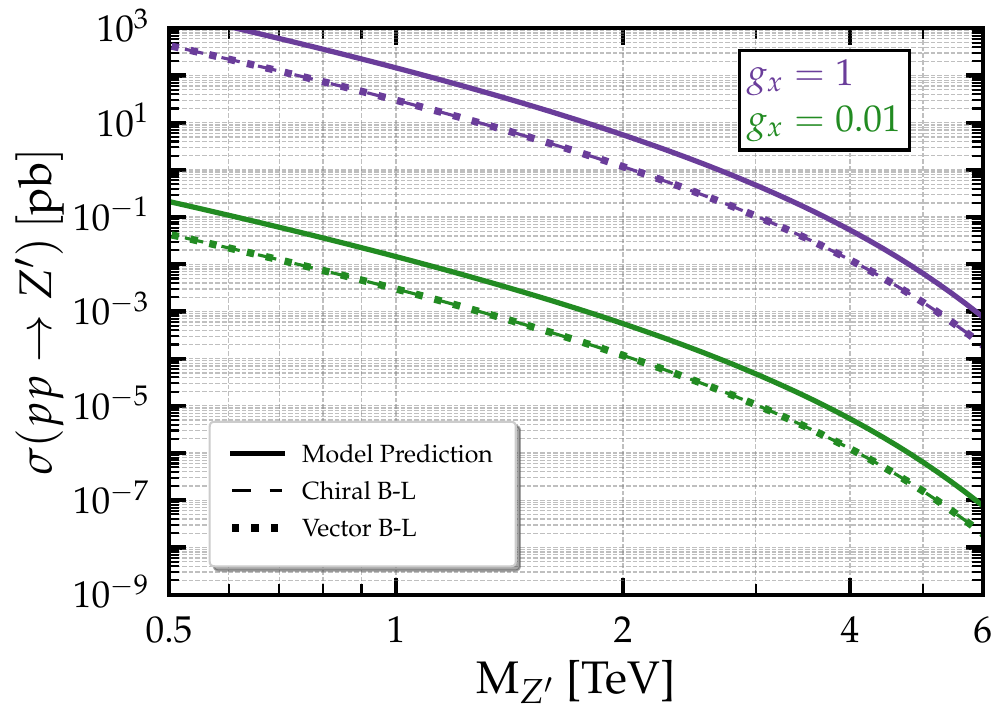}
    \caption{Production cross sections of the $Z'$ boson in $pp$ collisions at $\sqrt{s}=13$ TeV as a function of $M_{Z'}$. The green and purple curves correspond to the benchmark gauge couplings $g_x = 0.01$ and $g_x = 1$, respectively. For each benchmark, the solid, dashed, and dotted curves represent the predictions of the present model, the chiral $(B-L)$ model, and the vector $(B-L)$ model, respectively.}
     \label{fig:Sig_Zp}
 \end{figure}

We begin by briefly discussing the production of the $Z'$ boson at the LHC. Its production rate is determined by its interactions with the SM quarks and is therefore controlled by three key parameters: the $Z'$ mass $M_{Z'}$, the gauge coupling $g_x$, and the corresponding $U(1)_{X}$ charge assignments of the quarks.
To perform the collider analysis, we implement the model in the \textit{SARAH} package~\cite{Staub:2015kfa}, which is used to derive the complete particle spectrum, interaction vertices, and model files required for phenomenological studies. These are subsequently interfaced with \textit{SPheno}~\cite{Porod:2003um,Goodsell:2017pdq} to compute the mass spectrum and relevant decay widths. The resulting model files are then exported to \textit{MadGraph5}~\cite{Alwall:2014hca}, where the production cross sections are evaluated at leading order using proton-proton collisions with a center-of-mass energy of $\sqrt{s}=13$ TeV.
The corresponding production cross sections for the process $pp \to Z'$ are presented in Fig.~\ref{fig:Sig_Zp} as a function of the $Z'$ mass $M_{Z'}$.
To demonstrate the dependence on the gauge coupling, we consider two representative benchmark values, $g_x=1$ and $g_x=0.01$, shown by the purple and green curves, respectively.
Since the production cross section scales quadratically with the gauge coupling, $\sigma(pp \to Z^\prime) \propto g_x^2$, the corresponding cross section for any other value of $g_x$ can be readily obtained through a simple rescaling.
We further compare our results with the predictions of the chiral and vector $(B-L)$ models. The corresponding production cross sections are depicted by the dashed and dotted curves, respectively, using the same benchmark values of the gauge coupling $g_x$.  
Notably, the quark sector is identical in the chiral and vector $(B-L)$ models with respect to the $U(1)_{B-L}$ charge assignments. As a result, the production cross section for the $Z'$ boson is the same in both scenarios.
In the present model, $U(1)_{X}$ charge assignments for the quark sector are larger than those in the chiral and vector $(B-L)$ models, see Table \ref{Tab:U(1)_BM_Charges}. This enhances the $Z'$ couplings to quarks
and hence gives a larger production cross section at the LHC.
The results demonstrate the expected dependence of the production cross section on the model parameters: it decreases with increasing $M_{Z'}$, whereas larger values of $g_x$ yield systematically higher production rates over the full range of masses shown.

We now turn to the decay modes of the $Z'$ boson. In all three models
discussed above, the $Z'$ has two-body decays into the SM fermions and into
the right-handed neutrinos. In the present model, the $Z-Z'$ mixing
induces the additional channels $Z'\to W^{+}W^{-}$ and $Z'\to ZH_{1,2}$,
which are suppressed by the mixing angle $\alpha$. The width is therefore
dominated by decays into fermions, and the ratio of the total width to the
$Z'$ mass can be approximated as \footnote{For simplicity we assume
throughout this section that $Z'\to\chi_d\chi_d^{\ast}$ is kinematically
forbidden, so that the branching fractions quoted are the maximum values
attainable for the SM modes.}
\begin{equation}\label{Eq:DecayW/ZP}
    \frac{\Gamma_{Z'}}{M_{Z'}} \simeq \sum_{\psi}\frac{\Gamma\,(Z' \to \, \bar{\psi} \psi)}{M_{Z'}} = \frac{g_{x}^{2}}{48\pi}
    \sum_{\psi} N_{c}^{\psi}
    \left[\left(Q_{\psi_{L}}+Q_{\psi_{R}}\right)^{2}
    +\left(Q_{\psi_{L}}-Q_{\psi_{R}}\right)^{2}\right]\,.
\end{equation}
Here $\Gamma\!\left(Z'\to\bar{\psi}\psi\right)$ is the partial decay width into the fermion pair $\bar{\psi}\psi$, and the sum runs over all fermions. The color factor $N_{c}^{\psi}$ equals $3$ for quarks and $1$ for leptons, while $Q_{\psi_{L}}$ and $Q_{\psi_{R}}$ denote the $U(1)_X$ charges
of the left and right-handed components of $\psi$.

 \begin{figure}[!h]
    \centering    
    \includegraphics[width=0.6\linewidth]{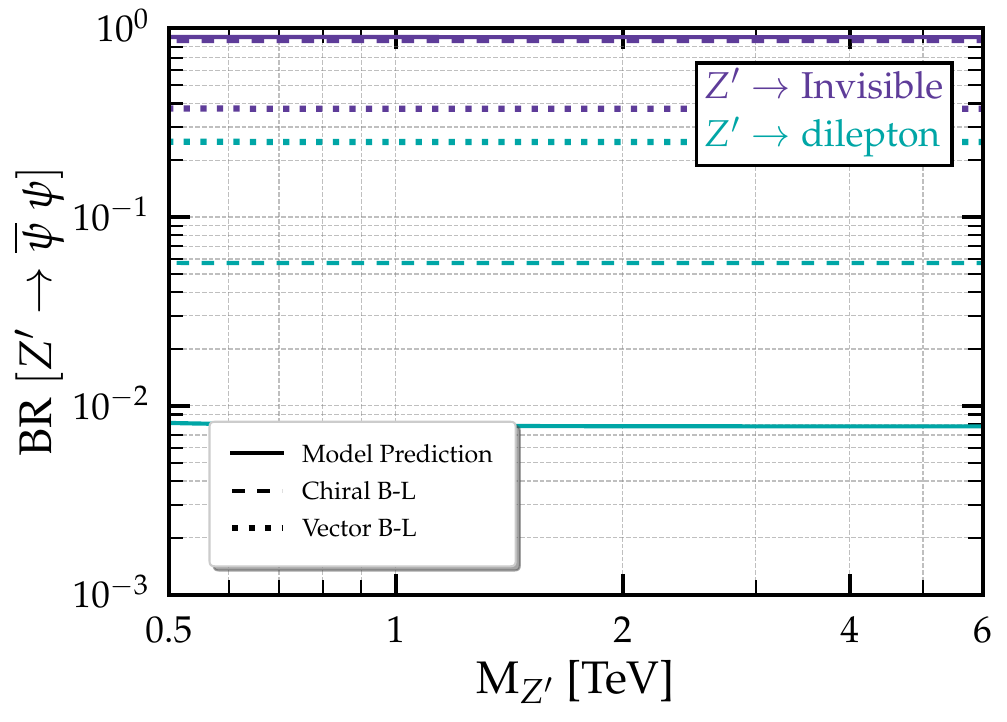}
     \caption{\centering Branching fractions of the $Z'$ boson as a function of its mass $M_{Z'}$.}
     \label{fig:Br_Zp1}
 \end{figure}

In Fig.~\ref{fig:Br_Zp1} we show the branching fractions of the $ Z' $ into dileptons and invisible decay channels.
The visible channels include quarks and charged leptons, while the invisible channels correspond to the neutrinos.
In plotting Fig. \ref{fig:Br_Zp1}, we assume that decays into DM are kinematically forbidden. The plotted branching fractions therefore represent the maximum possible values for the SM decay modes.
As shown in Fig.~\ref{fig:Br_Zp1}, the invisible channel dominates in the
present model, accounting for about $90\%$ of the total width. The
corresponding fraction is $86\%$ for the chiral $(B-L)$ case and $38\%$ for
the vector $(B-L)$ model. This behavior follows from the sizable $U(1)_X$ charges of the right-handed neutrinos, which give the largest single contribution to the total width.
The dilepton branching fraction shows the opposite ordering:
${\rm BR}(Z'\to\ell^{+}\ell^{-})$ is largest for the vector $(B-L)$
model at about $25\%$, falls to roughly $6\%$ in the chiral $(B-L)$ case, and
reaches only $0.8\%$ in the present model. Two effects contribute here. The
enlarged total width dilutes all branching fractions, and in addition $e_{R}$ is neutral under $U(1)_X$ for the benchmark charges of
Table~\ref{Tab:U(1)_BM_Charges}, so that the $Z'$ couples to charged leptons
only through the left-handed doublet.
It should be noted that if the decay $Z'\to\chi_{d}\chi_{d}^{\ast}$ is kinematically allowed, the branching fraction into dileptons will be
further reduced.

 \begin{figure}[!h]
     \centering      
     \includegraphics[width=0.6\linewidth]{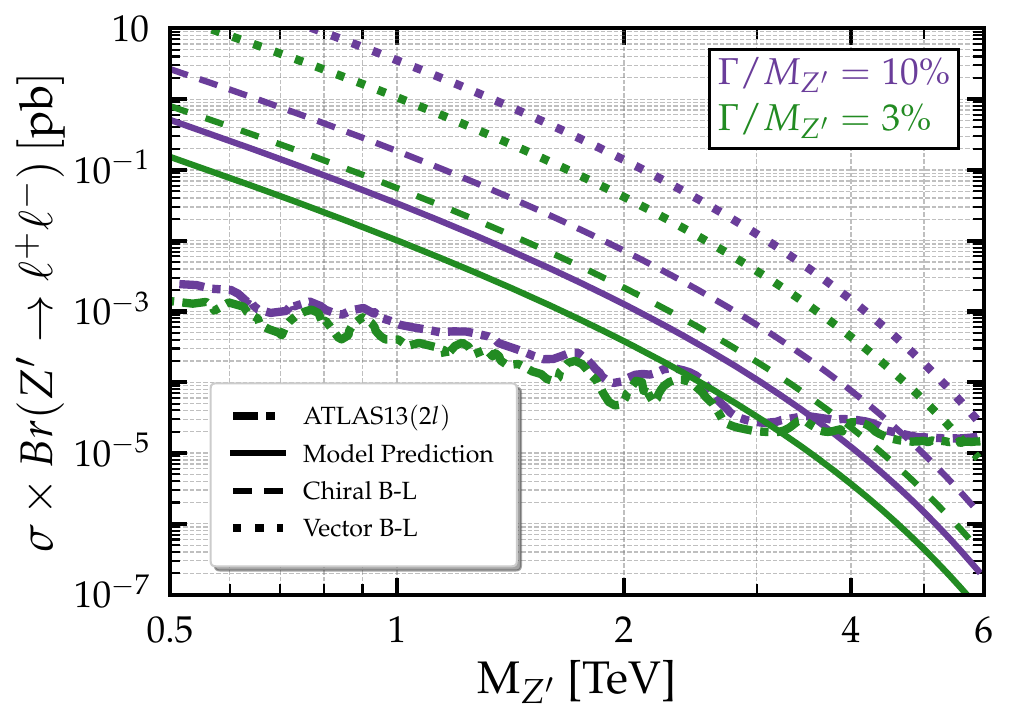}
\caption{Predicted $\sigma(pp \to Z')\times{\rm BR}(Z'\to\ell^{+}\ell^{-})$ as a function of $M_{Z'}$ for the model considered in this work (solid), the chiral $(B-L)$ case (dashed), and the vector $(B-L)$ case (dotted). Purple and green curves correspond to $\Gamma_{Z'}/M_{Z'}=10\%$ and $3\%$, respectively, corresponding to gauge couplings $g_x \simeq 0.17$ $(0.09)$ for the present model, $0.33$ $(0.18)$ for the chiral $(B-L)$ model, and $0.69$ $(0.38)$ for the vector $(B-L)$ model. The thick dash-dotted curves show the corresponding ATLAS $95\%$ C.L. upper limits at $\sqrt{s}=13$~TeV with $139~\text{fb}^{-1}$~\cite{ATLAS:2019erb}.}
\label{fig:Br_Zp2}
 \end{figure}

Combining the production cross section with the dilepton branching fraction, we can derive constraints on the $Z'$ mass.
In Fig.~\ref{fig:Br_Zp2}, we compare our model predictions with the ATLAS search for dilepton resonances using $pp$ collision data at $\sqrt{s} = 13$~TeV with an integrated luminosity of $139~\text{fb}^{-1}$~\cite{ATLAS:2019erb}, for two benchmark values of the relative width, $\Gamma_{Z'}/M_{Z'} = 10\%$ and $3\%$.
From Eq.~\eqref{Eq:DecayW/ZP}, the relative width is $\Gamma_{Z'}/M_{Z'} \simeq 0.21\,g_x^2$, $0.93\,g_x^2$, and $3.4\,g_x^2$ for the vector $(B-L)$, chiral $(B-L)$, and present model, respectively. Each choice of $\Gamma_{Z'}/M_{Z'}$ therefore fixes the gauge coupling: $\Gamma_{Z'}/M_{Z'} = 10\%$ $(3\%)$ corresponds to $g_x \simeq 0.69$ $(0.38)$ for the vector $(B-L)$ model, $g_x \simeq 0.33$ $(0.18)$ for the chiral $(B-L)$ model, and $g_x \simeq 0.17$ $(0.09)$ for the model considered here.
The dash-dotted purple and green curves
give the ATLAS limits for $\Gamma_{Z'}/M_{Z'}=10\%$ and $3\%$ respectively,
while the solid, dashed and dotted curves show the predictions of the
present model, the chiral $(B-L)$ model and the vector $(B-L)$ model, with the same color coding for the two width choices.
Comparing these predictions with the ATLAS limits gives a lower bound on the
$Z'$ mass of $3.5$~($3.2$)~TeV for $\Gamma_{Z'}/M_{Z'}=10\%$~($3\%$) in the
present model, and $4.6$~($3.9$)~TeV in the chiral $(B-L)$ case. Both are
considerably weaker than the corresponding bounds of $6.0$ and $5.6$~TeV obtained for the vector $(B-L)$ scenario. This relaxation originates from
the large $U(1)_X$ charges carried by the right-handed neutrinos, which enhance the invisible width and thereby suppress
${\rm BR}(Z'\to\ell^{+}\ell^{-})$ from $25\%$ in the vector case to $0.8\%$ for the present model.
The bound on $M_{Z'}$ is correspondingly relaxed by up to $\sim 2.5$~TeV relative to the vector $(B-L)$
model.
We note that other choices of $X_L$ and $\kappa$ modify the quark, charged lepton, and right-handed neutrino charges simultaneously. While larger right-handed neutrino charges suppress BR$(Z'\to\ell^+\ell^-)$, the accompanying increase of the quark charges enhances the $Z'$ production rate, so that the relaxation of the dilepton bound does not grow indefinitely with $\kappa$. Keeping $X_L=-1$ fixed, the benchmark $\kappa=2$ represents the minimal extension beyond the chiral $(B-L)$ model.

\begin{figure}[!h]
     \centering      
     \includegraphics[width=0.6\linewidth]{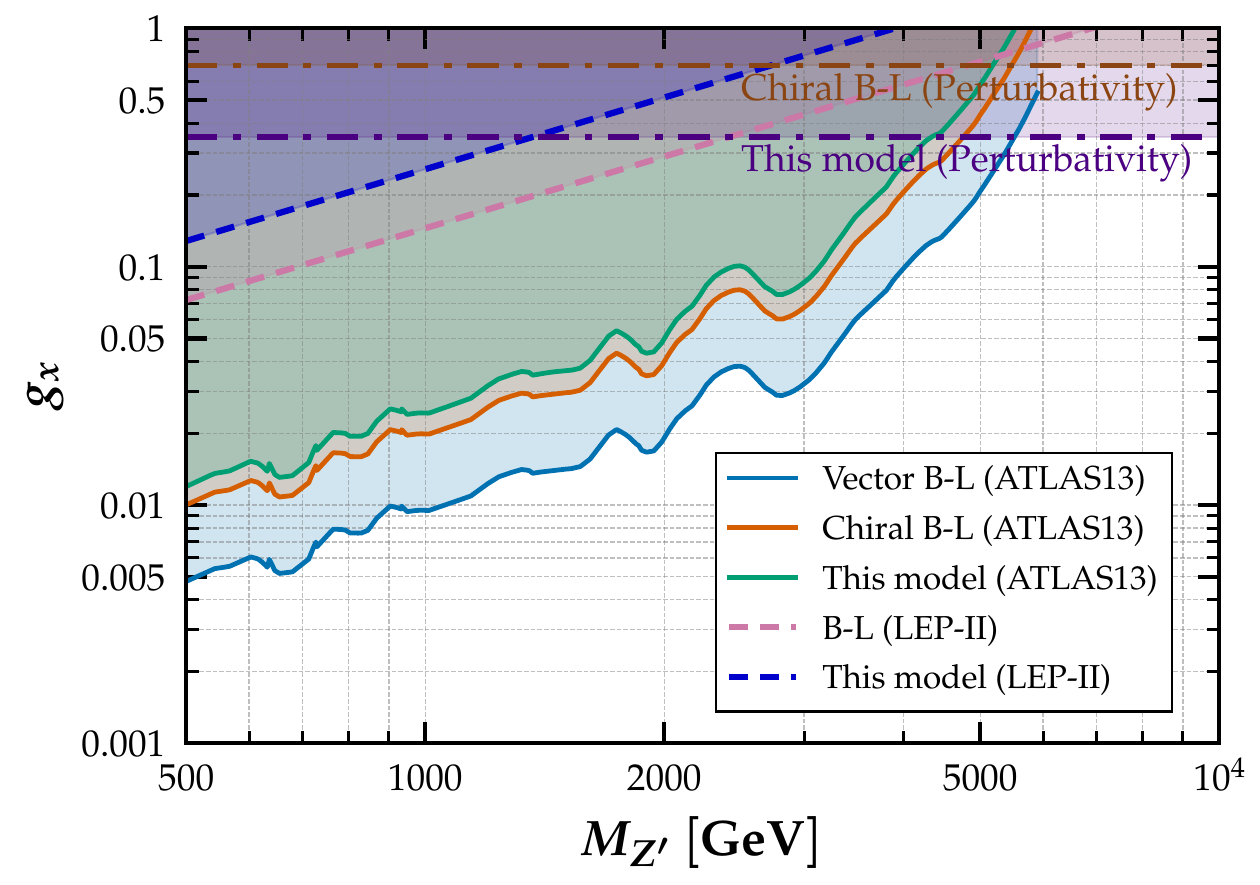}
\caption{Excluded regions in the $M_{Z'}-g_{x}$ plane. The solid curves
show the bounds from the ATLAS dilepton resonance search at
$\sqrt{s}=13$~TeV with $139$~fb$^{-1}$~\cite{ATLAS:2019erb} for the
vector $(B-L)$ model (blue), the chiral $(B-L)$ model (orange) and the
model considered in this work (green). 
The dashed lines give the
corresponding LEP-II contact interaction limits for both $(B-L)$ cases (pink) and the model presented here (blue dashed). The horizontal dash-dotted lines show the perturbativity constraints on the models. }
\label{fig:Zp_gx_Lim}
 \end{figure}

These limits can also be projected onto the $M_{Z'}-g_{x}$ plane. Following the approach of
Refs.~\cite{Das:2021esm,Prajapati:2024wuu,Kang:2026lgr}, we project these
limits onto the $M_{Z'}-g_{x}$ plane for the three charge assignments of
Table~\ref{Tab:U(1)_BM_Charges}. The resulting exclusion plot is shown in
Fig.~\ref{fig:Zp_gx_Lim}.
For each charge assignment, the shaded region above the corresponding curve is excluded, since the signal rate grows as
$g_{x}^{2}$ and eventually exceeds the measured upper limit.
The three exclusions are ordered according to their dileptonic branching
fractions, which are about $25\%$, $6\%$ and $0.8\%$ for the vector $(B-L)$ model, the chiral $(B-L)$ case, and the present model, respectively. 
The strongest bound is therefore obtained for the vector $(B-L)$ scenario and the weakest for the present model. 
At a given $M_{Z'}$, the bound on the gauge coupling is weakened by a factor of about $\sim 2.6$ for the model considered here, relative to the vector $(B-L)$ case.
Note that these bounds correspond to the conservative case in which the $Z'$ decays only into fermions. The inclusion of the
$Z'\to\chi_{d}\chi_{d}^{\ast}$ channel would further increase the total
width and reduce the dileptonic branching fraction, relaxing these mass limits still further.

Having discussed the constraints from hadron colliders, we now turn to those
from lepton colliders~\cite{Electroweak:2003ram,ALEPH:2013dgf}.\footnote{Where
the LHC searches apply, that is for $M_{Z'}\lesssim6$~TeV, they are generally
stronger than the LEP-II
bounds~\cite{Okada:2018ktp,Das:2021esm,Mandal:2023oyh,Prajapati:2026tfv}.}
LEP-II measured the cross section for $e^{+}e^{-}\to\bar{\psi}\psi$, where $\psi$ denotes the SM fermions, and placed limits on contact interactions of the form
$(\bar{e}\gamma^{\mu}P_{A}e)(\bar{\psi}\gamma_{\mu}P_{B}\psi)$ obtained by integrating out a heavy $Z'$ with $M_{Z'}\gg\sqrt{s}$. Here $P_{A,B}$ are chiral projection operators taking values $L$ or $R$.
They parametrize these effective contact
interactions
as~\cite{Electroweak:2003ram,Carena:2004xs,ALEPH:2013dgf,Huang:2019obt}
\begin{equation}\label{Eq:LEP_EFT}
    \mathcal{L}_{\rm eff} = \pm \frac{4\pi}
{\left(1+\delta_{e\psi}\right)\left(\Lambda_{AB}^{\psi\pm}\right)^{2}}
    \left(\overline{e}\gamma^{\mu}P_{A}e\right)
    \left(\overline{\psi}\gamma_{\mu}P_{B}\psi\right)\,,
\end{equation}
 where $\Lambda_{AB}^{\psi\pm}$ is the energy scale of the contact interaction and $\delta_{e\psi}$ is one for $\psi=e$ and $0$ otherwise.
For $M_{Z'}^{2}\gg s$, with $\sqrt{s}=209$~GeV at LEP-II, the exchange of a
heavy $Z'$ generates the effective interaction
\begin{equation}\label{Eq:LEP_EFT_Zp}
    \frac{1}{M_{Z'}^{2}}
    \left[\overline{e}\gamma^{\mu}
    \left(g^{z'}_{e_{L}}P_{L}+g^{z'}_{e_{R}}P_{R}\right)e\right]
    \left[\overline{\psi}\gamma_{\mu}
    \left(g^{z'}_{\psi_{L}}P_{L}+g^{z'}_{\psi_{R}}P_{R}\right)\psi\right]\,,
\end{equation}
where $g^{z'}_{\psi_{L(R)}}$ denotes the coupling of the $Z'$ to the fermion $\psi_{L(R)}$. Comparing with Eq.~\eqref{Eq:LEP_EFT} gives the
following bounds,
\begin{equation}
    M_{Z'}^{2} \geq \frac{1}{4 \pi}\, |g_{e_{_{A}}}^{z'}\, g_{\psi_{_{B}}}^{z'}|\,(\Lambda_{AB}^{\psi \pm})^{2}\,,
\end{equation}
where the two signs correspond to constructive and destructive interference
with the SM amplitude, so that $\Lambda_{AB}^{\psi+}$ applies for
$g^{z'}_{e_{A}}g^{z'}_{\psi_{B}}>0$ and $\Lambda_{AB}^{\psi-}$ for $g^{z'}_{e_{A}}g^{z'}_{\psi_{B}}<0$.
There are six chirality structures possible
$AB=\{LL,\,RR,\,LR,\,RL,\,VV,\,AA\}$, the last two denoting the purely
vector and axial-vector combinations. 
We take the measured bounds on the
$\Lambda_{AB}^{\psi\pm}$ from Ref.~\cite{ALEPH:2013dgf}.
These constraints depend only on the
$U(1)_X$ charges of the electron and of the final state fermion, and are
therefore fixed once the charge assignment is specified, independently of
the remaining details of the model \cite{Carena:2004xs}.
Which structure gives the strongest constraint depends on the charge assignment. For the $(B-L)$ model, the electron couples with equal strength to both chiralities, $X_{e_{L}}=X_{e_{R}}=-1$, and the interaction is purely vectorial; the corresponding limit
$\Lambda_{VV}^{\ell+}>24.6$~TeV translates into $M_{Z'}/g_{x}>6.9$~TeV. In the present model, by contrast, the right-handed charged leptons are neutral under $U(1)_X$, so that $g^{z'}_{e_{R}}=0$ and every structure involving a
right-handed electron vanishes identically.
The resulting bound is correspondingly
weaker, $M_{Z'}/g_{x}>3.9$~TeV.
For comparison with the ATLAS bounds, these limits are shown by the dashed lines in Fig.~\ref{fig:Zp_gx_Lim}.
A single line applies to both the vector and the chiral $(B-L)$ models, because the two assignments have identical lepton charges, $X_{L}=X_{e_{R}}=-1$.
Thus, both classes of collider constraints are relaxed in the present model, though for different reasons. The enlarged $U(1)_X$ charges of the right-handed neutrinos increase the invisible width and suppress ${\rm BR}(Z'\to\ell^{+}\ell^{-})$, weakening the ATLAS bound on $g_{x}$ by a factor of about $\sim 2.6$ relative to the vector $(B-L)$ case, while the vanishing right-handed electron charge removes the dominant contact operators and relaxes the LEP-II limit from $M_{Z'}>6.9\,g_{x}$~TeV to $M_{Z'}>3.9\,g_{x}$~TeV. 
While the large right-handed neutrino charges relax the dilepton
resonance bounds, they simultaneously tighten the perturbativity requirement. Fig.~\ref{fig:Zp_gx_Lim} therefore also shows the constraint
$X_{\nu_{R_3}}g_{x}\leq\sqrt{4\pi}$, indicated by the horizontal
dash-dotted lines for the chiral $(B-L)$ model and for the model
considered here. The resulting bound is stronger in the present case,
$g_{x}\lesssim0.35$, compared with $g_{x}\lesssim0.70$ for the chiral
$(B-L)$ model. At large $M_{Z'}$ values, where the dilepton searches
lose sensitivity, the present model is therefore more strongly
constrained by perturbativity.
Also, the parameter space opened up by the weaker LEP-II bound is
already excluded by the perturbativity requirement
$X_{\nu_{R_3}}g_{x}\leq\sqrt{4\pi}$, which restricts
$g_{x}\lesssim0.35$ in this model. Thus, the relaxation of the LEP-II
limit does not translate into additional viable parameter space, and
the net gain over the vector $(B-L)$ case comes entirely from the
weakened dilepton resonance bound.

We further note that the relatively large $U(1)_X$ charges, particularly
$X_{\nu_{R_3}}=10$ for our benchmark, can enhance the running of $g_x$
and lead to a Landau pole at high scales. At one loop
~\cite{Machacek:1983tz,Luo:2002ey}, neglecting kinetic mixing, the gauge
coupling evolves as $\beta(g_x)=b\,g_x^3/(16\pi^2)$ with $b\simeq196$,
giving
\begin{equation}
\Lambda_{\rm LP}\simeq\mu_0
\exp\left[\frac{8\pi^2}{b\,g_x^2(\mu_0)}\right].
\end{equation}
For $g_x\lesssim0.1$ and $\mu_0\sim M_{Z'}$ at the TeV scale, the
formal one-loop Landau pole lies above the Planck scale. For larger
couplings, the Landau pole occurs below the Planck scale, and the model
should therefore be regarded as an effective theory valid below the
corresponding cutoff. 
For comparison, the same one-loop estimate gives $b\simeq12$
and $54$ for the vector and chiral $(B-L)$ models, respectively, for which the Landau pole lies above
the Planck scale for $g_x\lesssim0.43$ and $0.20$.

Another constraint associated with such relatively large $U(1)_X$
charges can arise from the thermal history of the Universe. The
right-handed neutrinos can interact with the thermal plasma
through the $Z'$ boson and may therefore be thermalized in the early
Universe. This can lead to additional cosmological constraints on
$M_{Z'}/g_x$, depending on the decoupling temperature \cite{Anchordoqui:2014dpa,Planck:2018vyg}. A quantitative
determination requires a dedicated analysis of the early Universe
thermal history and the corresponding Boltzmann equations, which is
beyond the scope of the present work. We leave a detailed study of
these cosmological constraints, including the associated $N_{\rm eff}$
implications for future work.

%

\section{Dark Sector Phenomenology} \label{sec:dm}

We now turn to a detailed discussion of the dark sector phenomenology of the model. The scalar field $\chi_{d}$ serves as the DM candidate, which is an SM singlet but carries charge $q_{\rm DM}$ under $U(1)_X$.
Being an SM singlet, it interacts with the visible sector only through the extended gauge and scalar sectors. The $U(1)_X$ charge assignment of $\chi_{d}$ is chosen such that all renormalizable operators leading to its decay are forbidden, thereby ensuring its stability without the need to impose any additional discrete symmetry. In our present framework, we have taken $q_{\rm DM}=5/2$.
Furthermore, since $\chi_{d}$ is charged under the gauged $U(1)_X$ symmetry, the dark sector phenomenology is intimately connected to the properties of the corresponding $Z'$ gauge boson. Consequently, the viable DM parameter space is also subject to constraints arising from collider searches for the $Z'$ boson.
In addition, our model features an extended scalar sector. In the minimal singlet scalar DM scenario, the direct detection constraint is typically satisfied only in the narrow Higgs resonance region. In contrast, the extended gauge and scalar sectors in our framework open up additional annihilation channels, significantly enlarging the viable DM parameter space while remaining consistent with current direct detection bounds. 
In the previous section, we discussed the collider constraints on the $Z'$ boson in our framework and compared them with those in the vector and chiral $(B-L)$ scenarios for the right-handed neutrinos. 
We found that the collider bounds are considerably weaker in the present model due to the specific $U(1)_X$ charge assignments of the quarks and leptons.
Motivated by this, we first look at and analyze the contribution from the $ Z'$-mediated annihilation channels. 
We then extend the discussion to the complete framework by incorporating all relevant processes, highlighting the impact of the extended gauge and scalar sectors on the viable DM parameter space.


\subsection{DM parameter space with pure gauge interactions}

\begin{figure}[!h]
     \centering      
     \includegraphics[width=0.99\linewidth]{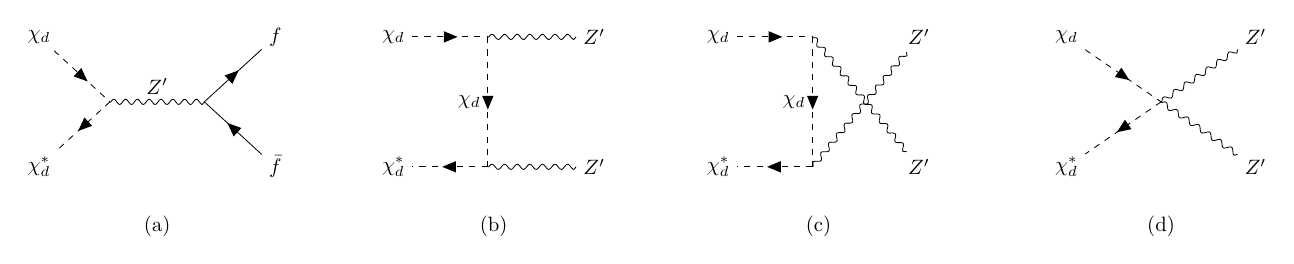}
\caption{Feynman diagrams for the dominant dark matter annihilation channels in the $Z'$ portal. Diagram (a) shows the $s$-channel
annihilation into fermions, $\chi_d\chi_d^{\ast}\to Z'^{\ast}\to f\bar{f}$,
where $f$ runs over the SM and BSM fermions. Diagrams (b), (c), and (d) show
the $t$-channel, $u$-channel, and contact contributions to
$\chi_d\chi_d^{\ast}\to Z'Z'$, which is open for $M_{\rm DM}>M_{Z'}$.}
     \label{Fig:Feynman}
 \end{figure}
\begin{figure}[!h]
    \centering
    \includegraphics[width=0.38\linewidth]{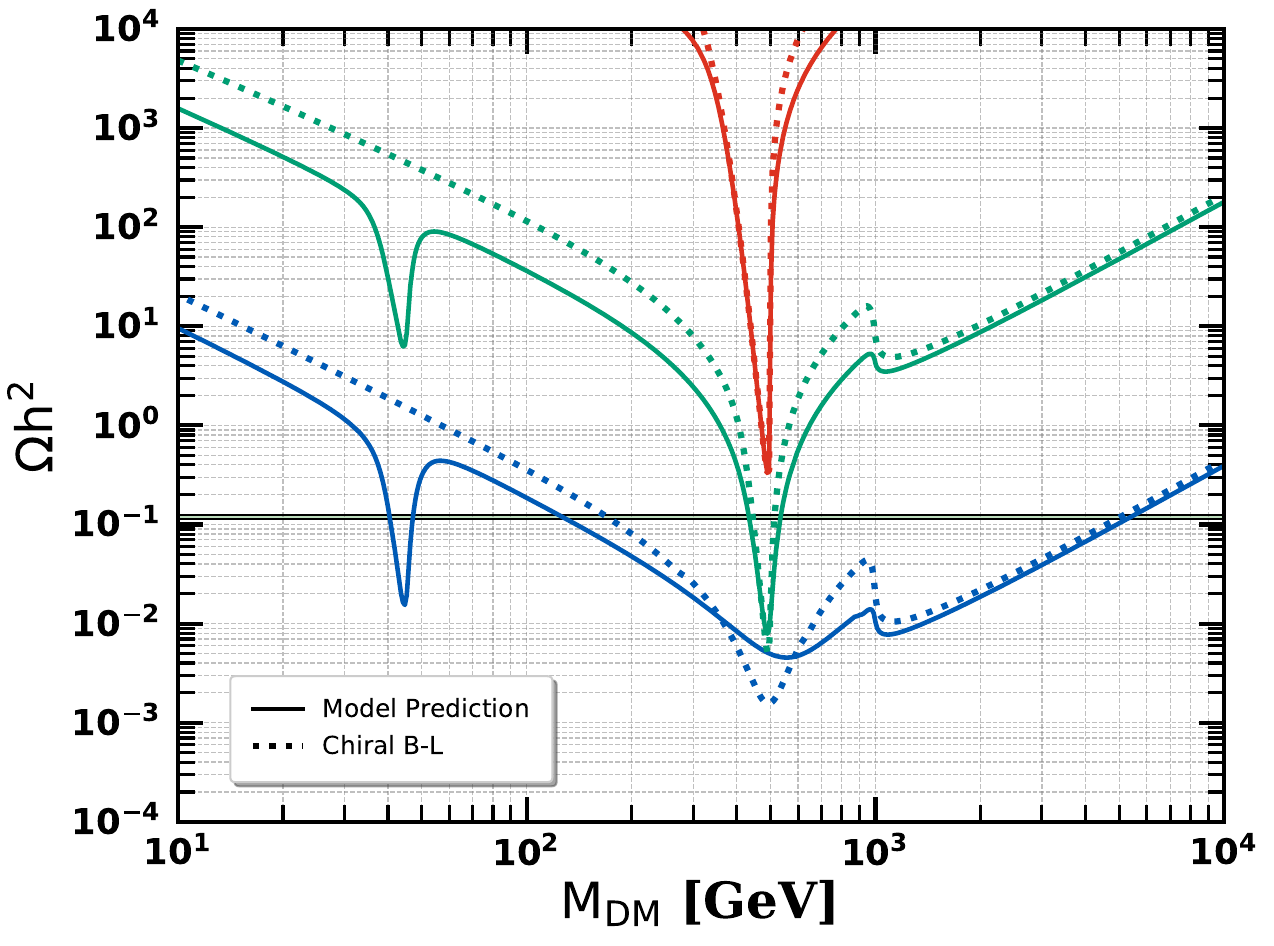}    \includegraphics[width=0.49\linewidth]{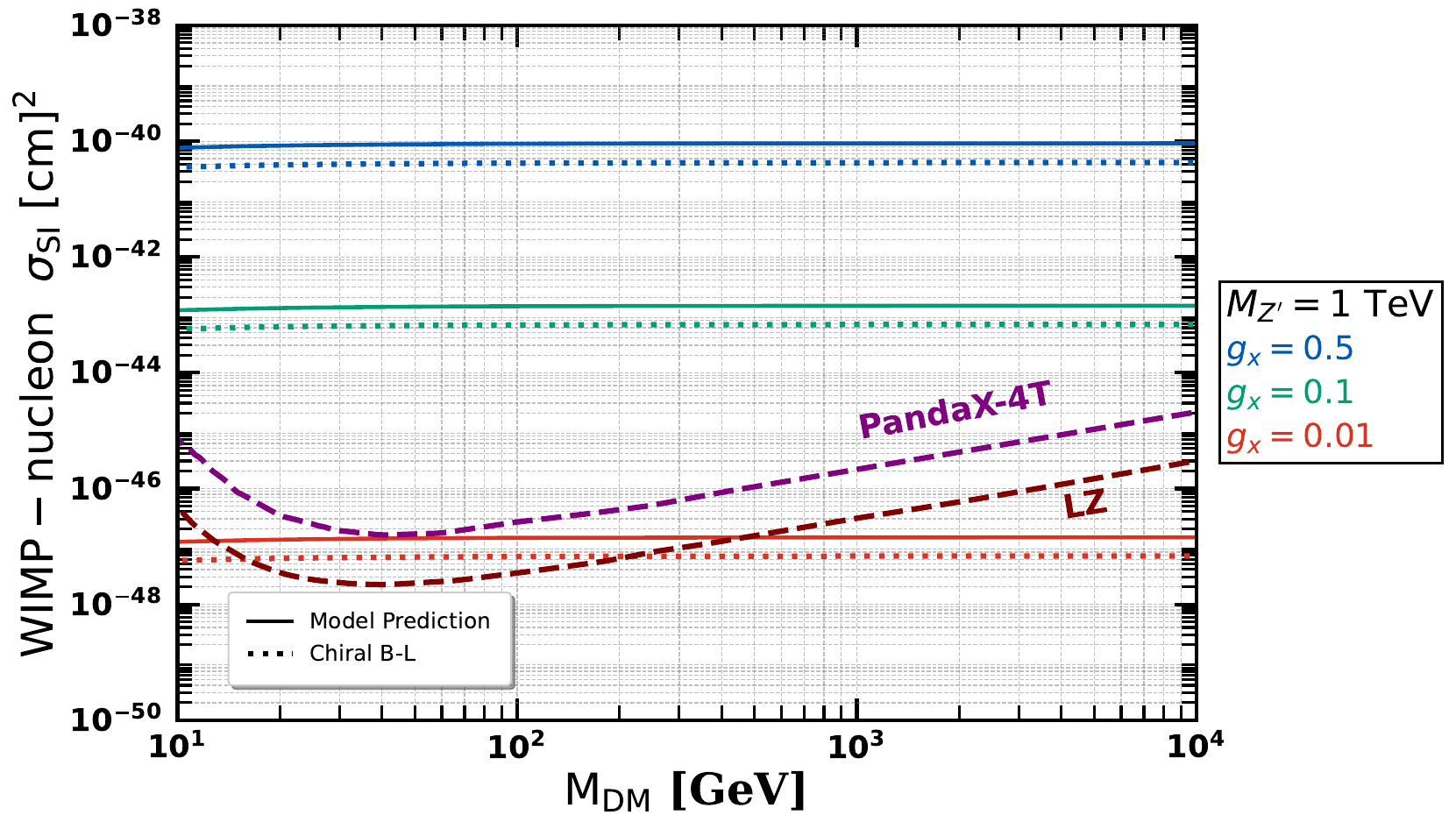}
    \includegraphics[width=0.38\linewidth]{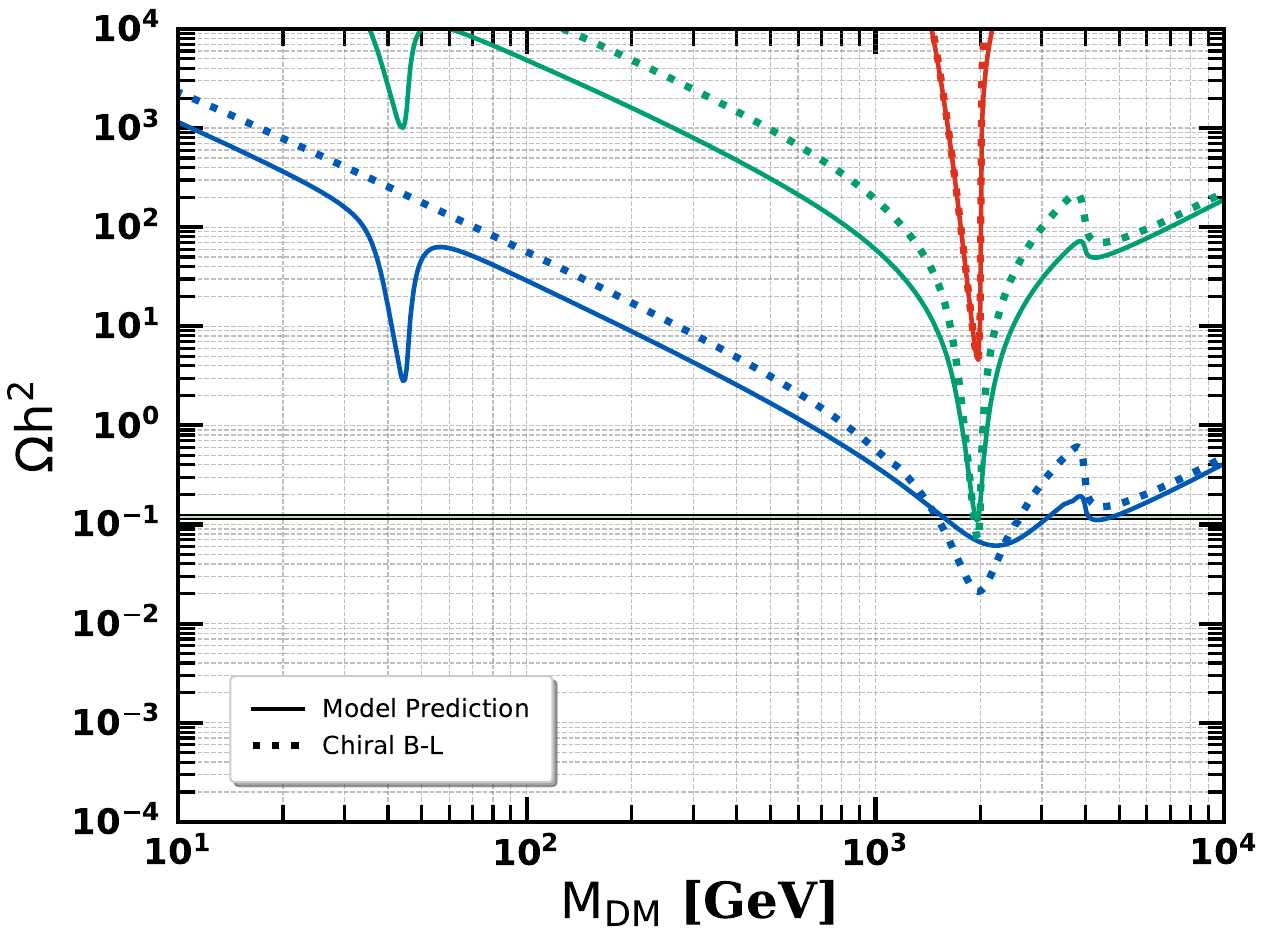}    \includegraphics[width=0.49\linewidth]{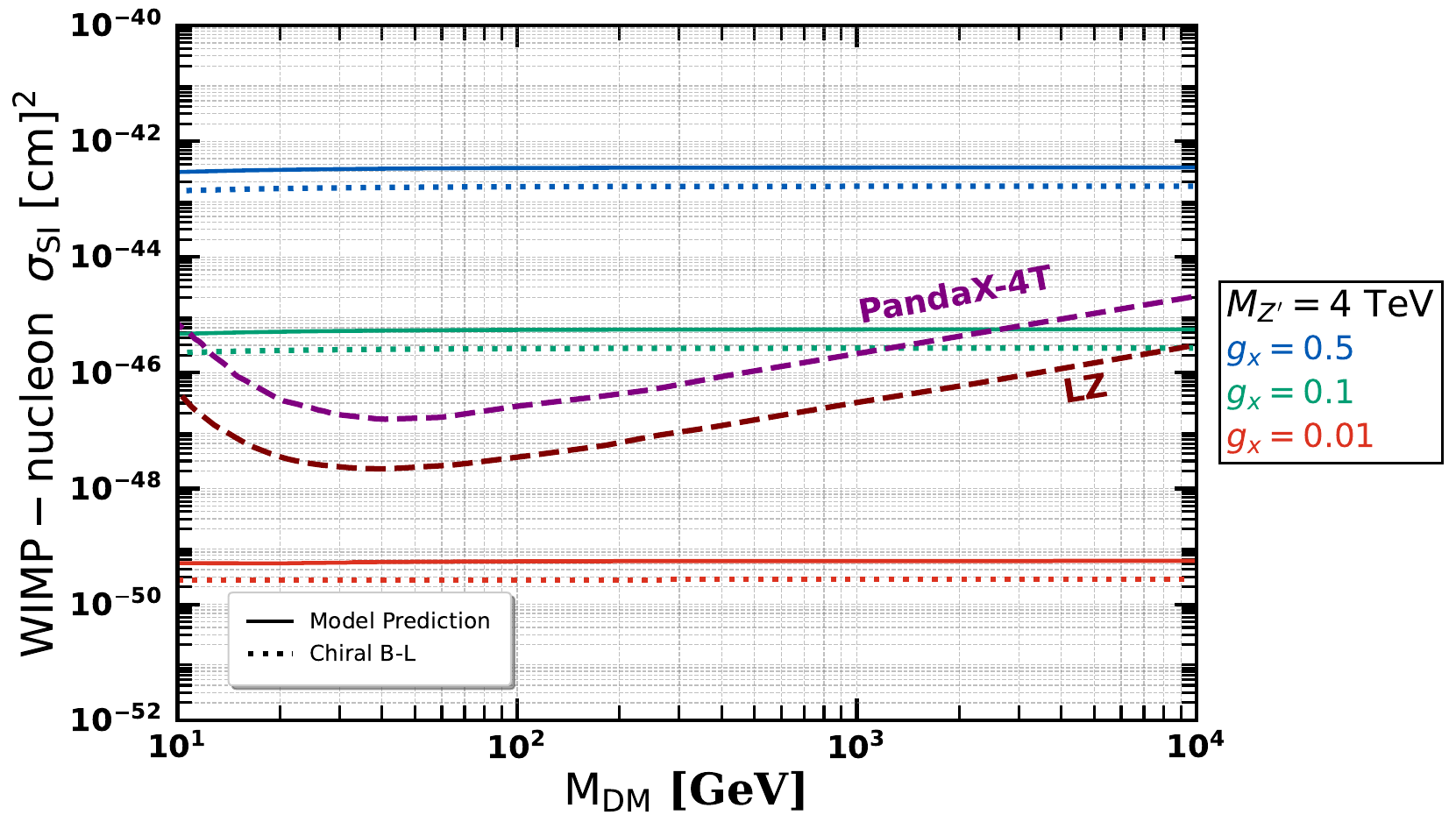}
    \includegraphics[width=0.38\linewidth]{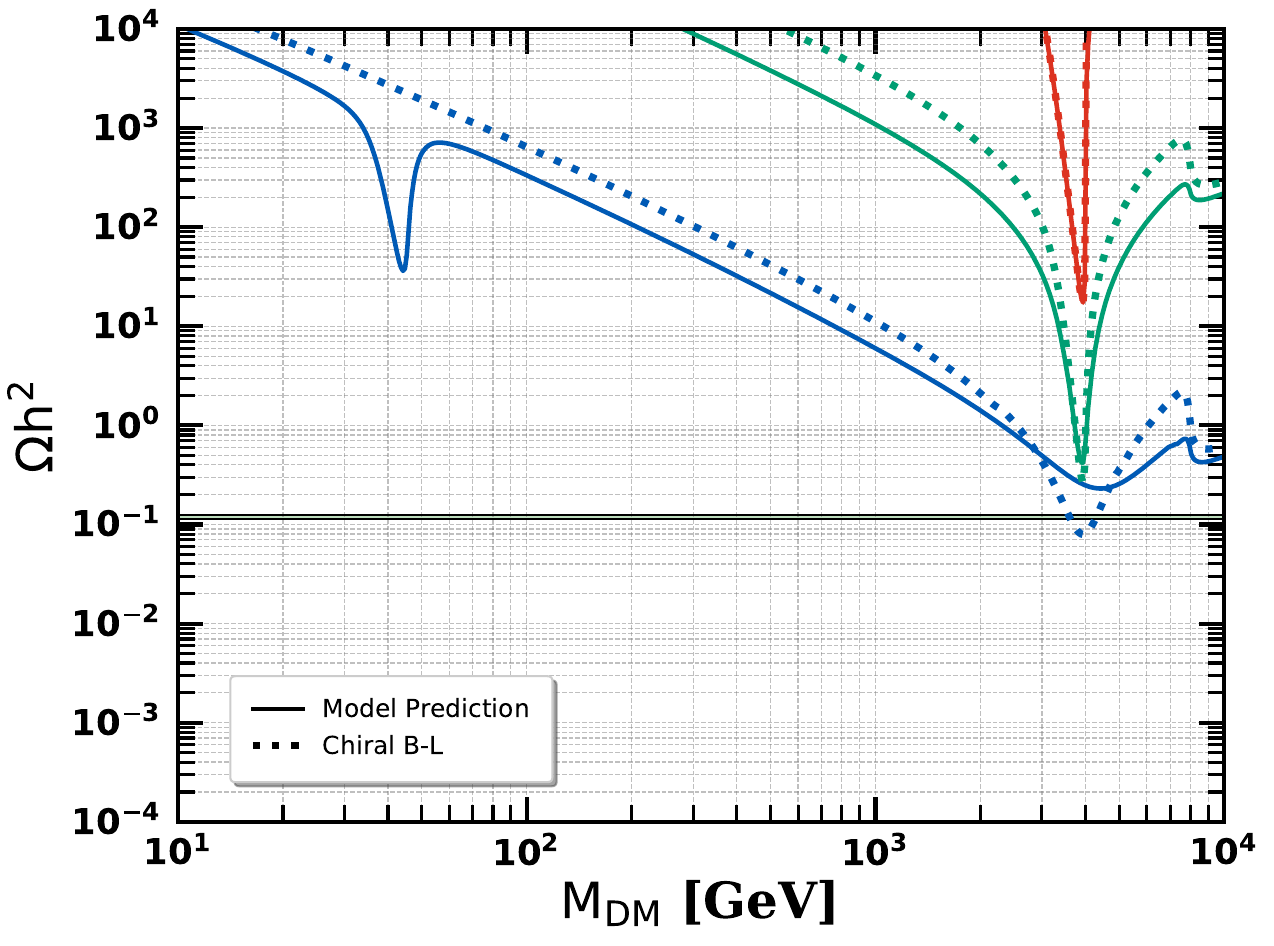}    \includegraphics[width=0.49\linewidth]{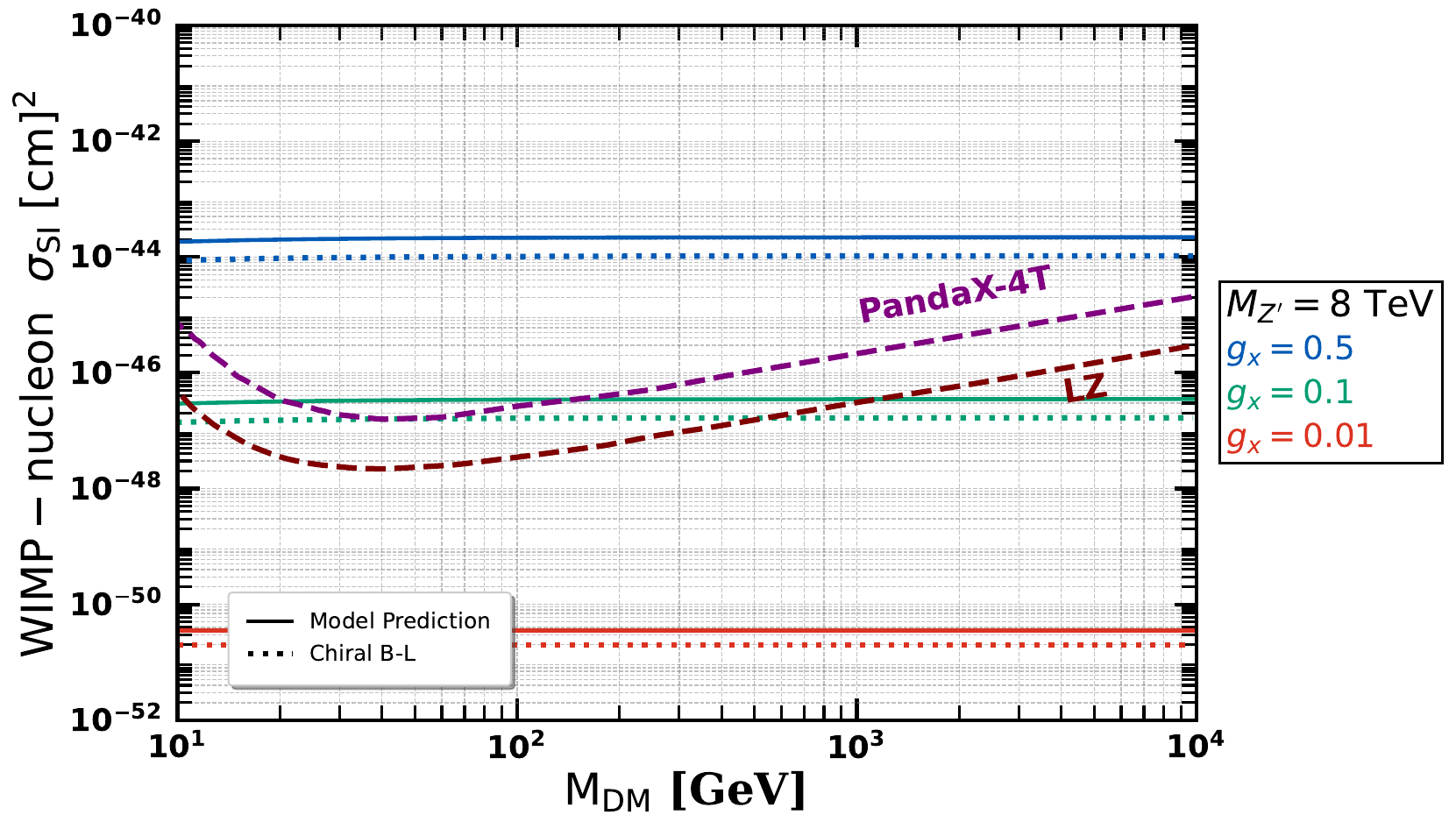}
    \caption{The DM parameter spaces for purely gauge interactions have been shown. The left and right panels illustrate the relic density and spin-independent WIMP-nucleon scattering cross section as a function of the DM mass $M_{\rm DM}$, respectively. The different rows correspond to $Z'$ masses $M_{Z'} = 1,\ 4,\ 8$ TeV (from top to bottom). In each panel, the colored curves represent different values of the gauge coupling $g_x$. The solid (dashed) curves depict our model (chiral $(B-L)$ model) predictions. The horizontal green band on the left panel shows the $3\sigma$ allowed range for DM \cite{Planck:2018vyg}.  The dashed purple and red lines in the right panels show the latest upper bounds from the PandaX-4T~\cite{PandaX:2024qfu} and LZ~\cite{LZ:2024zvo} collaborations, respectively.}
    \label{fig:OnlyZP_S1DM}
\end{figure}

As discussed above, the DM candidate $\chi_d$ carries a non-zero $U(1)_X$
charge and therefore interacts directly with the corresponding gauge boson
$Z'$. To disentangle the impact of the gauge sector from that of the
extended scalar sector, we first consider the limiting case in which the DM phenomenology is driven exclusively by $ Z'$-mediated interactions.
In this scenario, the DM mainly annihilates into fermions through the $s$-channel process
$\chi_d \chi_d^{\ast} \to Z'^{\ast} \to f \bar{f}$, where $f$ denotes all
the SM and BSM fermions.\footnote{When the Higgs is charged under $U(1)_X$,
$Z-Z'$ mixing is induced and additional channels open, namely
$\chi_d \chi_d^{\ast} \to Z^{\ast} \to f \bar{f}$, and
$\chi_d \chi_d^{\ast} \to W^{+} W^{-}$. However, these channels are suppressed by the
mixing angle.}
For $M_{\rm DM} > M_{Z'}$, the DM can also annihilate into a pair of gauge bosons, $\chi_d \chi_d^{\ast} \to Z'Z'$. The former process is $p$-wave and hence the resulting $\langle \sigma v \rangle$ is velocity suppressed, whereas the latter is not \cite{Rodejohann:2015lca,Berlin:2018sjs}.
The Feynman diagrams for these annihilation channels are shown in
Fig.~\ref{Fig:Feynman}.
The corresponding annihilation cross section is controlled primarily by the
$U(1)_{X}$ gauge coupling $g_{x}$, the $Z'$ and DM masses, and the
$U(1)_{X}$ charges of the DM and of the fermions.
Therefore, once the $U(1)_{X}$ charges are fixed, the DM phenomenology is
governed by only three independent parameters: the DM mass $M_{\rm DM}$,
the gauge coupling $g_{x}$, and the $Z'$ mass $M_{Z'}$.
We again choose the benchmark models given in Table \ref{Tab:U(1)_BM_Charges}.
The corresponding DM parameter spaces are illustrated in Fig. \ref{fig:OnlyZP_S1DM}. The left panel shows the relic density as a function of the DM mass for representative choices of the gauge coupling $g_x$. The horizontal green band indicates the $3\sigma$ range of the observed  DM relic abundance measured by Planck \cite{Planck:2018vyg}. 
Here the solid curves represent the predictions of our model, while the
dashed curves correspond to the chiral $(B-L)$ scenario for the same set of
benchmark parameters. The red, green and blue curves denote $g_{x}=0.01$,
$0.1$ and $0.5$ respectively. The first, second and third rows correspond
to $M_{Z'}=1$, $4$ and $8$ TeV respectively.
The right panels show the corresponding spin-independent WIMP-nucleon
scattering cross section $\sigma^{\rm SI}$ as a function of the DM mass
$M_{\rm DM}$, with the same benchmarks and color coding. The current
exclusion limits from PandaX-4T~\cite{PandaX:2024qfu} and LZ~\cite{LZ:2024zvo} are shown as the
dashed purple and dark red curves, respectively.
The gauge coupling $g_{x}$ and the mediator mass $M_{Z'}$ act in opposite
senses on $\Omega h^{2}$ and $\sigma^{\rm SI}$. A larger $g_{x}$ enhances
the annihilation cross section and thus reduces $\Omega h^{2}$, while
increasing $\sigma^{\rm SI}$. Increasing $M_{Z'}$ instead suppresses the
annihilation rate, giving a larger $\Omega h^{2}$ and a correspondingly
smaller $\sigma^{\rm SI}$. This is apparent from a comparison of the first
and third rows of Fig.~\ref{fig:OnlyZP_S1DM}.
For $g_{x}\lesssim0.1$ the annihilation rate is insufficient to deplete the
DM abundance and the relic density is over-abundant across most of the
parameter space. Larger couplings enhance the annihilation efficiency and
allow the observed abundance to be reproduced over a wider range of DM
masses, but they are strongly constrained by the collider bounds and perturbativity constraints discussed
in Sec.~\ref{sec:Zprime}. The associated predictions for $\sigma^{\rm SI}$
also exceed the current direct detection limits over a substantial region of
parameter space, as shown in the right panels of
Fig.~\ref{fig:OnlyZP_S1DM}.
A similar comparison between the chiral and vector $(B-L)$ models is presented in Appendix~\ref{sec:Appen1}.

For the relic density plots in the left panels of
Fig.~\ref{fig:OnlyZP_S1DM}, both the solid and dashed curves exhibit a
pronounced dip in the vicinity of the $Z'$ resonance,
$M_{\rm DM}\sim M_{Z'}/2$, where the resonantly enhanced annihilation cross
section strongly depletes the DM abundance. A second, comparatively milder
suppression appears at $M_{\rm DM}\sim M_{Z'}$, once the $Z'Z'$ final state
becomes kinematically accessible through
$\chi_d\chi_d^{\ast}\to Z'Z'$.
A further feature is present at $M_{\rm DM}\sim M_{Z}/2$, arising from
annihilation through the SM $Z$ pole. This is absent in the chiral $(B-L)$
framework, where the SM Higgs is uncharged under the gauge symmetry and no
$Z-Z'$ mixing is generated. In our model, the non-zero $U(1)_X$ charge of
the Higgs induces such mixing after electroweak symmetry breaking, opening
an additional resonant annihilation channel.

Apart from the $Z$ pole, the two benchmarks can be compared in three
distinct regimes, separated by the $Z'$ resonance at $M_{\rm DM} = M_{Z'}/2$ and by
the opening of the $\chi_d \chi_d^{*} \to Z'Z'$ channel at $M_{\rm DM} = M_{Z'}$.
Below the resonance, for $M_{\rm DM} \ll M_{Z'}/2$,
the $Z'$ is off shell and annihilation proceeds entirely through
$\chi_d \chi_d^{*} \to Z'^{*} \to f\bar{f}$. Owing to the large $U(1)_X$
charges of the right-handed neutrinos, the annihilation cross section is
larger in our model than in the chiral $(B-L)$ case, and the relic
density is correspondingly lower. The two curves remain parallel, as both
follow the same scaling $\Omega h^{2} \propto M_{Z'}^{4} /(g_{x}^{2} M_{\rm DM})^{2}$.
Near the resonance, in the vicinity of $M_{\rm DM} \approx M_{Z'}/2$, the annihilation cross section is given by the Breit-Wigner profile \cite{Gondolo:1990dk,Griest:1990kh,Ibe:2008ye,Guo:2009aj,Duch:2017nbe}.
The shape of the dip is controlled by the ratio of the fractional width
$\Gamma_{Z'}/M_{Z'}$ to the thermal spread in the center-of-mass energy at
freeze-out, $1/x_f \approx 0.04$. When $\Gamma_{Z'}/M_{Z'} \ll 1/x_f$, the
resonance is narrower than the thermal distribution and the total width
cancels, leaving approximately
$\left(\Omega h^{2}\right)_{\rm min} \propto M_{Z'}^{2}/ \left(q_{\rm DM} g_x\right)^{2}$ \cite{Ibe:2008ye,Guo:2009aj}.
Since $g_x$ and $q_{\rm DM}$ are common to both models, the two dips coincide, as
observed at the smaller values of the gauge coupling.
When instead $\Gamma_{Z'}/M_{Z'} \gg 1/x_f$, the resonance is broader than
the thermal distribution and approximately
$\left(\Omega h^{2}\right)_{\rm min} \propto M_{Z'}^{2} \Gamma_{Z'}/\Gamma_{Z' \to \chi_{d} \chi_{d}^{*}}$. Increasing the
width then merely lowers the peak of the Breit-Wigner profile, so that the
dip becomes simultaneously shallower and broader. This is the behavior seen
in our model relative to the chiral $(B-L)$ case \cite{Ibe:2008ye,Guo:2009aj}. 
We emphasize that the
same enhancement of $\Gamma_{Z'}$ responsible for the loss of depth is
precisely what suppresses ${\rm BR}(Z' \to \ell^{+}\ell^{-})$ and relaxes
the ATLAS dilepton constraint; the relief obtained at the LHC is paid for at
the resonance.
Above the $Z'$ mass, once $M_{\rm DM} > M_{Z'}$, the channel $\chi_d \chi_d^{*} \to Z'Z'$ opens, producing the kink visible in Fig.~\ref{fig:OnlyZP_S1DM}. This process is mediated entirely by the DM-$Z'$ vertex and therefore depends only on $g_x$ and $q_{\rm DM}$. The two
curves overlap in this regime because
$\chi_d \chi_d^{*} \to Z'^{*} \to f\bar{f}$ is velocity suppressed whereas
$\chi_d \chi_d^{*} \to Z'Z'$ is not, so that the latter dominates at high
mass.
Taken together, the enlarged right-handed neutrino charges act in opposite
senses on the two constraints. The enhanced total width suppresses
${\rm BR}(Z' \to \ell^{+}\ell^{-})$ and thereby relaxes the ATLAS dilepton
bound, but it simultaneously raises the minimum of the resonant dip. As is
evident from Fig.~\ref{fig:OnlyZP_S1DM}, the relic abundance and direct detection
constraints can be satisfied simultaneously only in the vicinity of the
resonance in $Z'$-mediated scenarios. Increasing the total width, therefore, narrows the viable DM parameter space.

From this discussion and plots, we can infer that the singlet DM candidate is strongly constrained in the pure $Z'$-portal scenario. Additional annihilation channels are therefore required to simultaneously satisfy the direct detection constraint and reproduce the observed relic abundance, as we discuss next. 


\subsection{Combined effect of gauge and scalar-mediated channels}

\begin{figure}[!h]
    \centering   \includegraphics[width=0.42\linewidth]{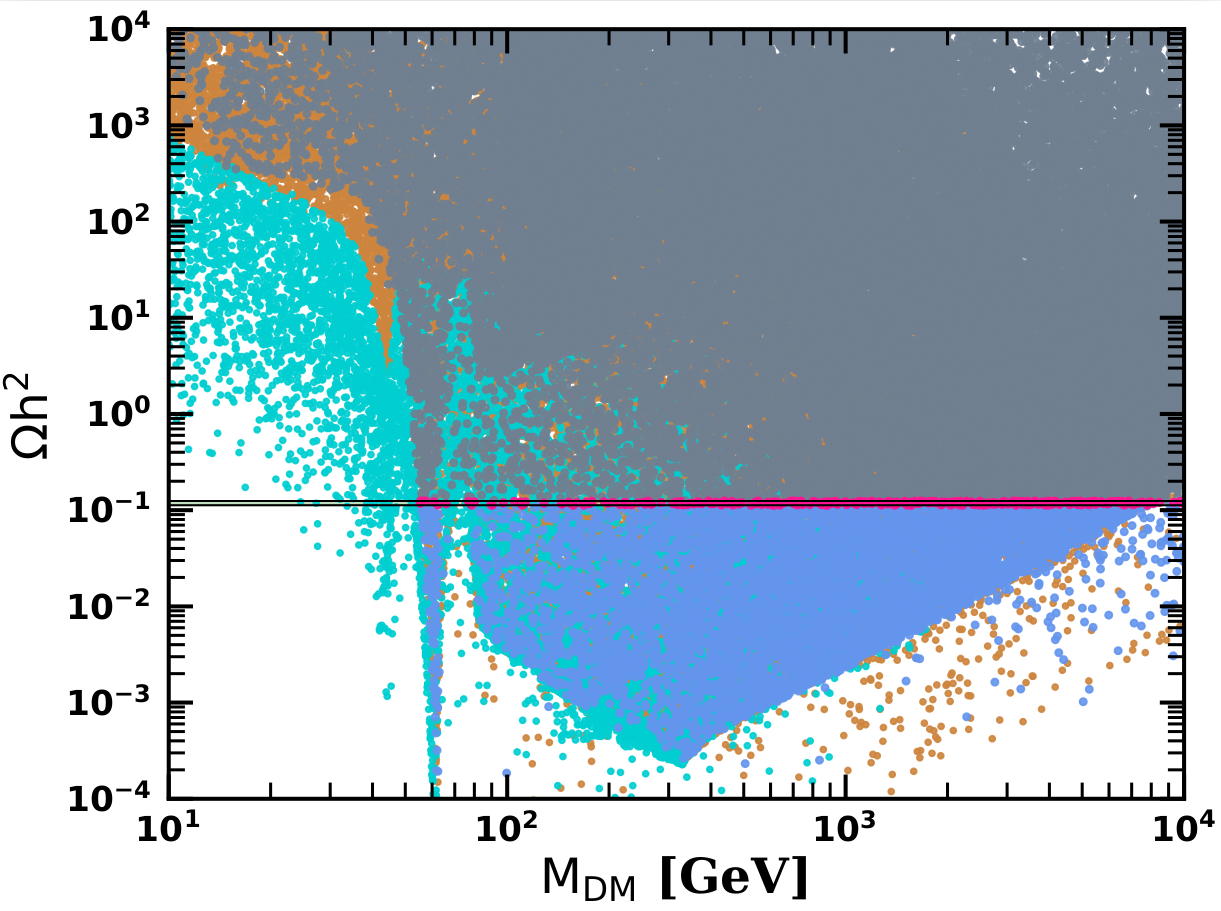}    \includegraphics[width=0.57\linewidth]{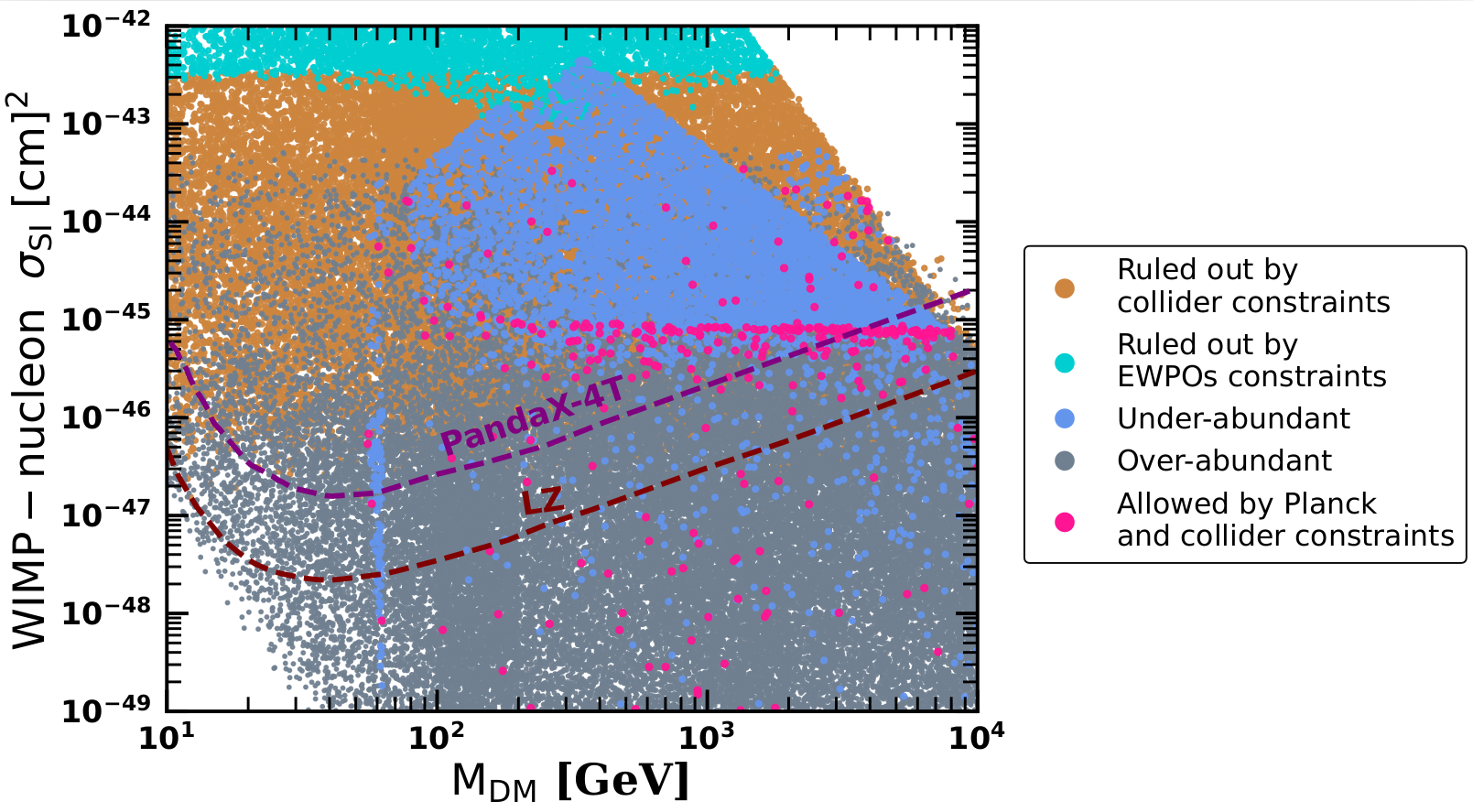}
    \caption{The DM parameter space including both gauge and scalar annihilation channels is shown. The left and right panels represent the relic density $\Omega h^2$ and spin-independent WIMP–nucleon cross section $\sigma^{\rm SI}$ as a function of the DM mass $M_{\rm DM}$, respectively. The orange points are excluded by collider bounds from LEP-II and LHC, and the cyan points are ruled out by electroweak precision constraints. The blue (gray) points correspond to under-abundant (over-abundant) relic density and the magenta points satisfy collider and electroweak precision constraints and lie within the $3\sigma$ allowed range for DM.}
 \label{fig:Relic_DD_Final}
\end{figure}

Having discussed the purely gauge-interaction case, we now turn to the full DM parameter space of our model, where both gauge and scalar-mediated interactions contribute to the DM annihilation processes. 
In contrast to the previous scenario, the DM phenomenology is no longer controlled solely by the $Z'$ portal. The extended scalar sector provides additional annihilation channels, thereby relaxing the strong correlation between the annihilation rate and the direct detection cross section that was present in the pure gauge case. In particular, the relic abundance can receive significant contributions from scalar-mediated processes even for relatively small values of the gauge coupling, allowing the direct detection rate to remain sufficiently suppressed.
In this case, the dominant annihilation channels are
$
\chi_d\chi_d^\ast \rightarrow Z'^\ast \rightarrow f\bar f,
$
together with the corresponding scalar-mediated processes,
$
\chi_d\chi_d^\ast \rightarrow H_i^\ast \rightarrow f\bar f,
$
and, whenever kinematically accessible, annihilation into scalar final states,
$
\chi_d\chi_d^\ast \rightarrow H_iH_j.
$
The relative importance of these channels depends on the DM mass, scalar and $Z'$ masses, as well as the corresponding gauge and scalar couplings. Consequently, the correct relic abundance need not be obtained only through the $Z'$ resonance, as was essentially required in the purely gauge-interaction scenario. The additional scalar interactions can instead provide an efficient annihilation mechanism over a much wider range of DM masses.
For this scenario, the gauge parameters are scanned over the ranges
$g_{x}\in[10^{-4},\,0.35]$ and $M_{Z'}\in[0.3,\,20]$~TeV, while the scalar
quartic couplings are varied in $[10^{-4},\,\sqrt{4 \pi}]$, subject to the theoretical
constraints discussed in Sec.~\ref{subsec:Scalar}.
The resulting parameter space is shown in Fig.~\ref{fig:Relic_DD_Final}, with the
left and right panels displaying $\Omega h^{2}$ and $\sigma^{\rm SI}$ as
functions of the DM mass $M_{\rm DM}$. The horizontal green band in the left panel indicates the $3\sigma$ range for the observed DM abundance,
$0.1164\le \Omega h^{2}\le0.1236$~\cite{Planck:2018vyg}. The latest exclusion limits from PandaX-4T~\cite{PandaX:2024qfu} and LZ~\cite{LZ:2024zvo} are shown by the purple and dark red curves, respectively.
The orange points are excluded by collider constraints. These comprise the ATLAS dilepton resonance searches discussed in Sec.~\ref{sec:Zprime}, together with the LEP-II bounds derived from searches for effective four-fermion interactions mediated by the $Z'$ \cite{ALEPH:2013dgf}, and the LHC constraints on invisible Higgs decays \cite{ATLAS:2023tkt}.
The cyan points are excluded by electroweak precision constraints, namely the $\rho$ parameter and the effective $Z$-boson couplings to charged leptons, which constrain the $Z-Z'$ mass mixing induced by the non-zero $U(1)_X$ charge of the Higgs doublet.
Among the points surviving these constraints, the blue and gray points correspond respectively to under-abundant and over-abundant relic densities.
The magenta points reproduce the observed relic abundance while satisfying
all of the above constraints. In the right panel, those magenta points lying
below the LZ limit constitute the viable parameter space of the model.

The surviving points of the model are found over the range $M_{\rm DM}\simeq60$~GeV to $10$~TeV. This is in contrast to the pure $Z'$-portal case and the minimal singlet scalar DM scenario, where simultaneously reproducing the observed relic abundance and satisfying direct-detection constraints restricts the viable DM mass range much more severely.

We emphasize that, over most of the viable parameter space, the relic abundance is determined predominantly by the scalar-mediated annihilation channels, while the $Z'$-mediated contributions are relevant mainly near the $Z'$ resonance. The $U(1)_X$ symmetry nevertheless remains essential for the DM sector: it guarantees the stability of $\chi_d$ without any ad hoc discrete symmetry, and the heavier CP-even scalar $H_2$, which mediates the dominant annihilation channels, is predominantly the $U(1)_X$ breaking singlet $\chi$. Moreover, the $Z'$ contribution to direct detection, together with the collider and electroweak constraints on $(M_{Z'},g_x)$, shapes the viable parameter space.

\section{Conclusion} \label{sec:conc}

In this work, we have proposed a simple generalization of the chiral $(B-L)$ framework through a non-universal charge assignment in the fermion sector. The primary motivation for this generalized charge structure is to suppress the dilepton branching fraction of the $Z'$, thereby substantially weakening the stringent collider constraints on the new gauge boson.
We extend the SM gauge symmetry by an additional $U(1)_X$ gauge symmetry. Gauge anomaly cancellation is ensured by introducing right-handed neutrinos $\nu_{R_i}$ ($i=1,2,3$) that are singlets under the SM gauge group but carry non-zero $U(1)_X$ charges. Solving the gauge anomaly cancellation conditions yields the general charge assignment for the fermion fields, with all charges parametrized by two independent parameters, $X_L$ and $\kappa$. The resulting $U(1)_X$ charges of $\nu_{R_i}$ are: ($-4 \kappa, -4 \kappa, 5\kappa$).

The $U(1)_X$ symmetry is spontaneously broken by the VEV of an additional scalar singlet $\chi$, which sets the $Z'$ gauge boson mass. Neutrino masses are generated through the VEV of $\chi$ via the effective operator
$
\overline{L}\tilde{\Phi}\nu_{R_\alpha}\chi$ $ (\alpha=1,2)
$.
The right-handed neutrino carrying the charge $5\kappa$ does not participate in neutrino mass generation, leaving one neutrino massless. We provide a UV completion of this effective operator by introducing two Dirac pairs of BSM fermions, $(N_L,N_R)$, through which neutrino masses arise via a Dirac type-I seesaw mechanism. Under the $U(1)_X$ symmetry, the SM Higgs carries the charge $(-X_L-\kappa)$, while the scalar $\chi$ and the BSM fermions $(N_L,N_R)$ carry charges $3\kappa$ and $-\kappa$, respectively.
To realize a sufficiently suppressed dilepton branching fraction of the $Z'$ and thereby relax the collider constraints, we consider the specific choice $X_L = -1,\, \kappa = 2$. This represents a minimal generalization of the chiral $(B-L)$ charge assignment, which is recovered for $X_L = -1,\, \kappa = 1$. 
We find that the generalized charge assignment substantially relaxes the collider constraints on the $Z'$ compared with the conventional vector and chiral $(B-L)$ scenarios.

We further investigate the DM phenomenology of the model by considering a singlet scalar DM candidate, $\chi_d$, interacting through both the additional gauge and scalar sectors. A notable feature of the framework is that no additional discrete symmetry is required to stabilize the DM candidate; its stability follows naturally from the chosen $U(1)_X$ charge assignment. The same charge structure that suppresses the dilepton branching fraction and alleviates the collider constraints, however, also has important consequences for the DM phenomenology. In particular, in the pure $Z'$-portal limit, the enhanced couplings to the fermion sector lead to a tension between achieving the observed relic abundance and satisfying the stringent direct detection constraints. The viable parameter space is therefore highly restricted when only gauge-mediated annihilation channels are considered.

The inclusion of the additional scalar-mediated annihilation channels resolves this tension by providing alternative mechanisms for efficiently depleting the DM abundance without requiring a large gauge coupling.
In this scenario, the relic abundance is predominantly set by the scalar-mediated channels, while the $U(1)_X$ symmetry guarantees the stability of the DM candidate and shapes the viable parameter space.
Consequently, the combined gauge and scalar-mediated scenario yields a substantially broader viable parameter space. We find that DM masses in the range $M_{\rm DM}\simeq 60~{\rm GeV}-10~{\rm TeV}$ can simultaneously satisfy the observed relic abundance, direct detection limits, and the relevant collider and electroweak precision constraints.
This broad mass range is in sharp contrast to the pure $Z'$-portal case and the minimal singlet scalar DM scenario, where the simultaneous requirements of obtaining the observed relic abundance and satisfying direct detection bounds severely restrict the viable parameter space.
Overall, our generalized chiral $U(1)_X$ framework provides a simple and economical realization in which the collider constraints on the $Z'$ are substantially relaxed through the generalized charge structure, while simultaneously accommodating neutrino masses and a viable DM candidate.

\section*{Acknowledgments}
\noindent
  R.K. is supported by the National Research Foundation of Korea under Grant NRF-2023R1A2C100609111. The work of H.K.P. is supported by the Prime Minister Research Fellowship (ID: 0401969). 
\FloatBarrier
\appendix

\section{Comparison of the chiral and vector $(B-L)$ models}
\label{sec:Appen1}
\begin{figure}[!h]
    \centering
    \includegraphics[width=0.38\linewidth]{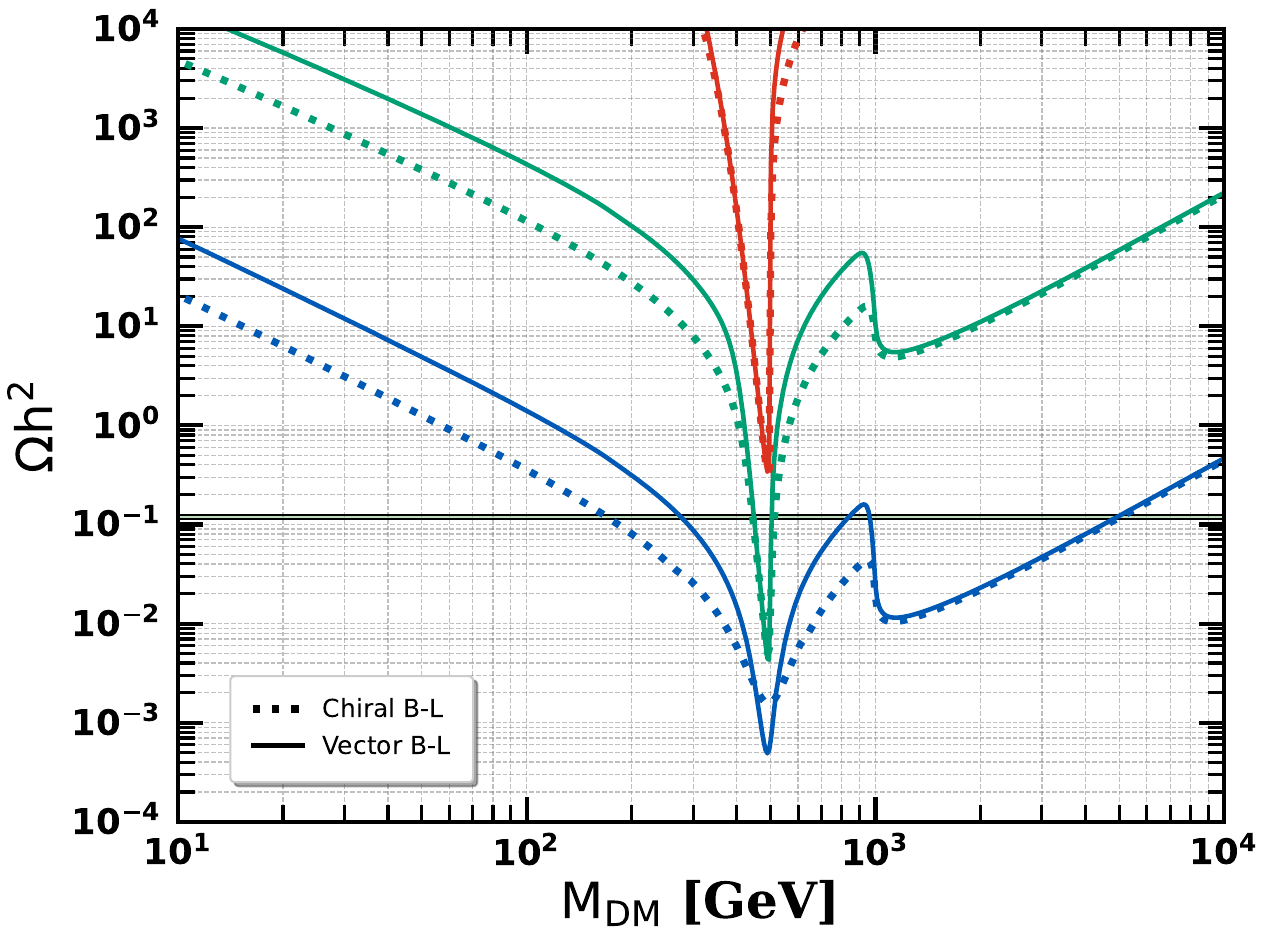}    \includegraphics[width=0.49\linewidth]{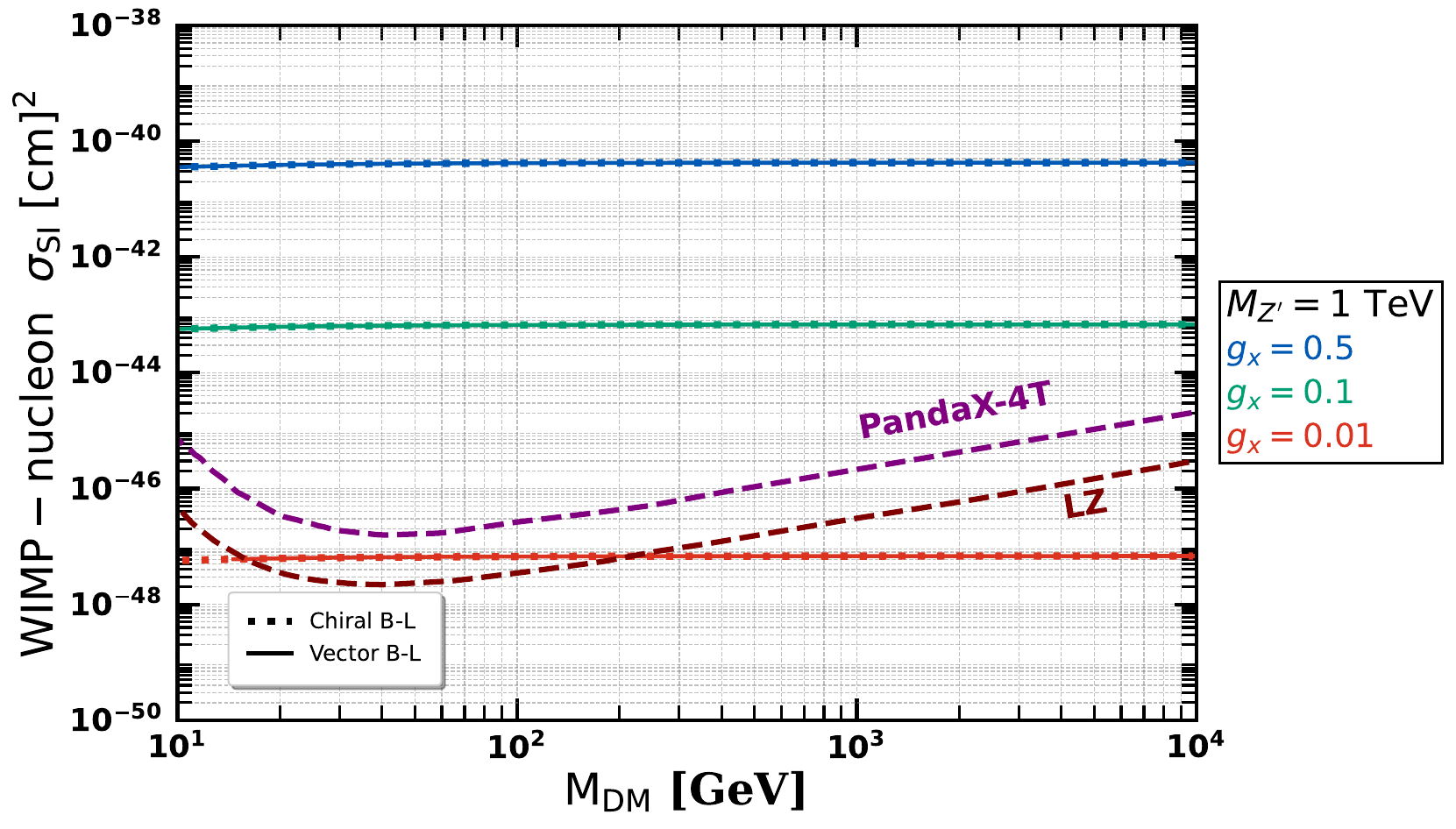}
    \includegraphics[width=0.38\linewidth]{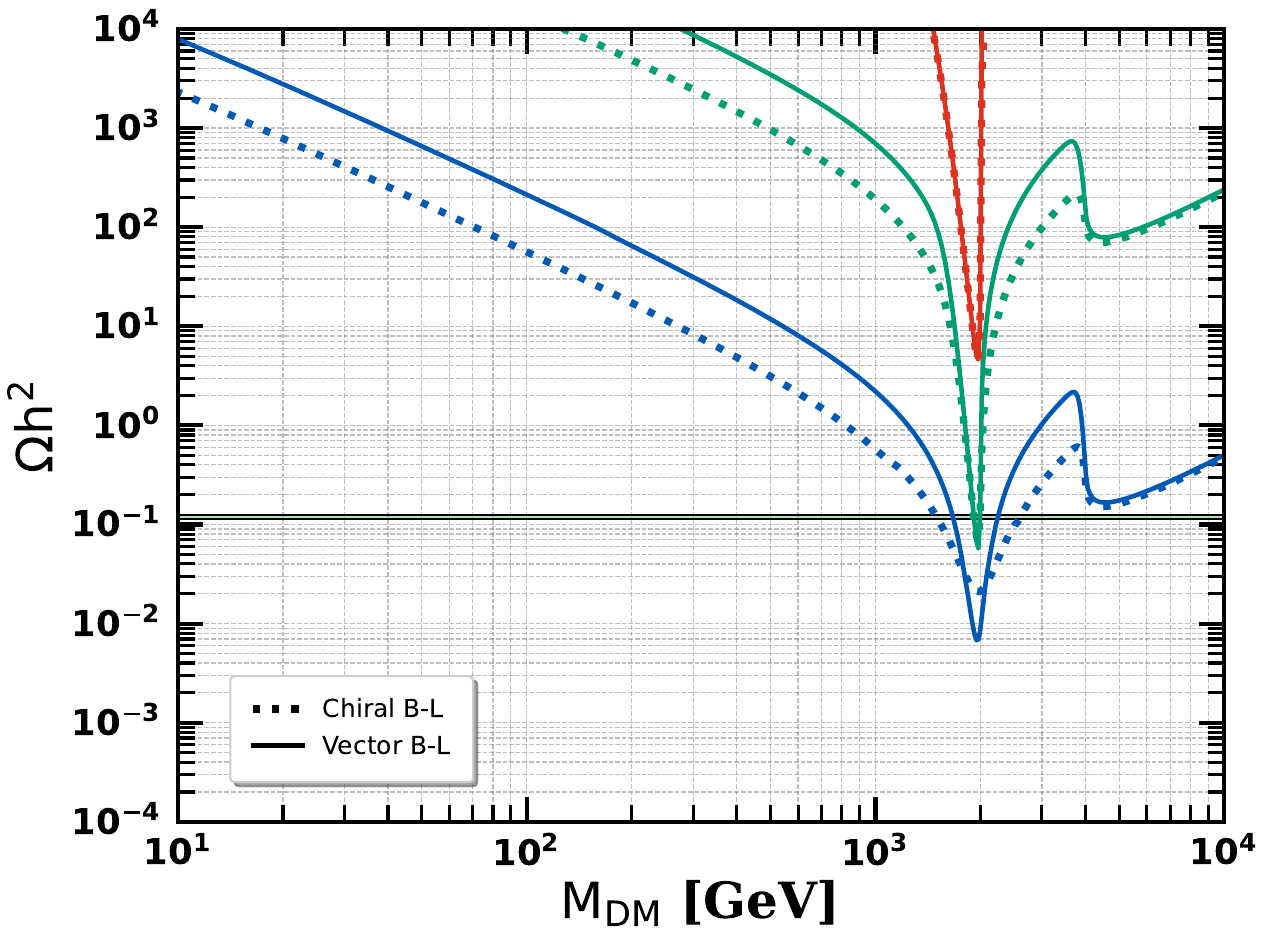}    \includegraphics[width=0.49\linewidth]{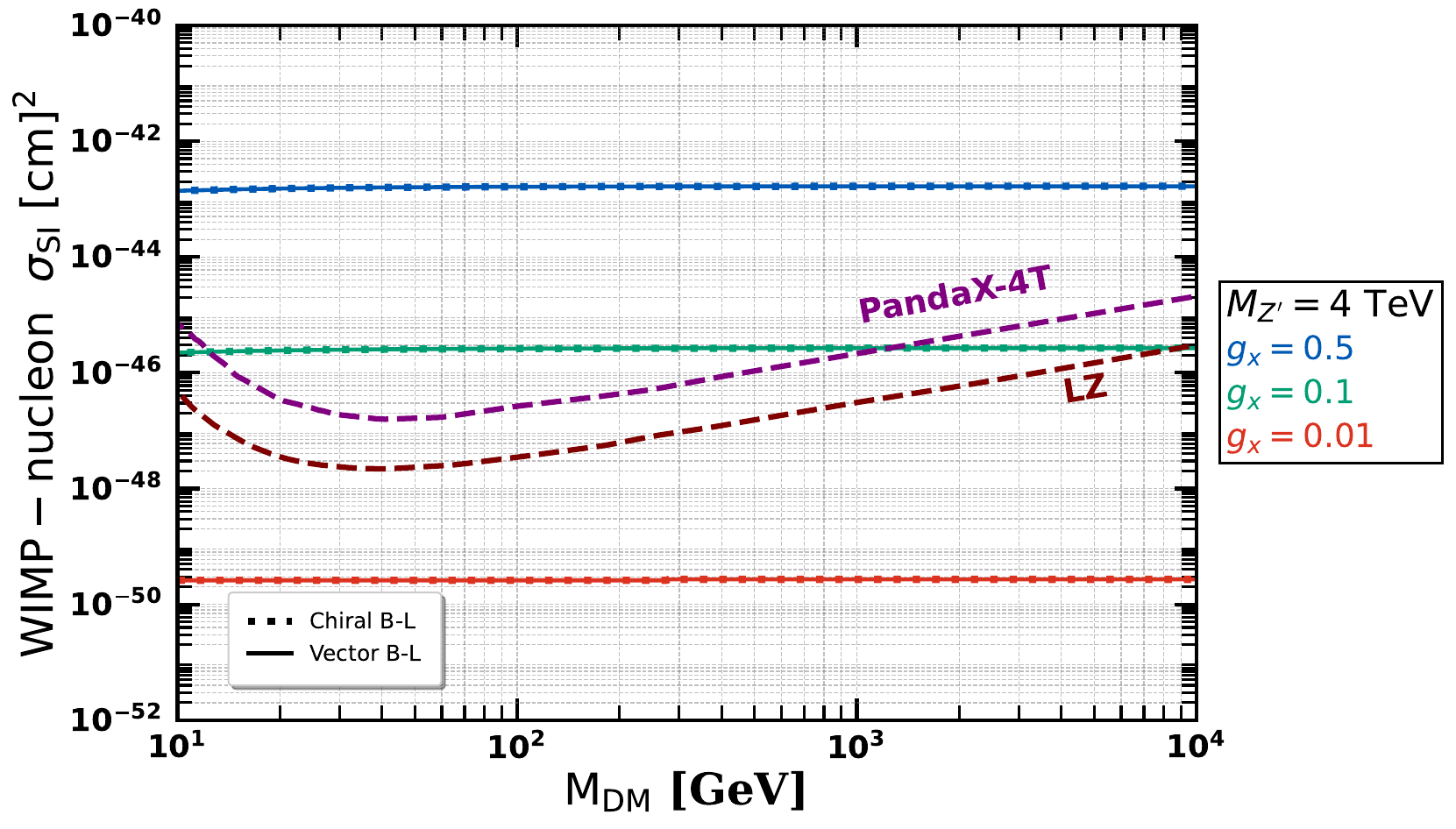}
    \includegraphics[width=0.38\linewidth]{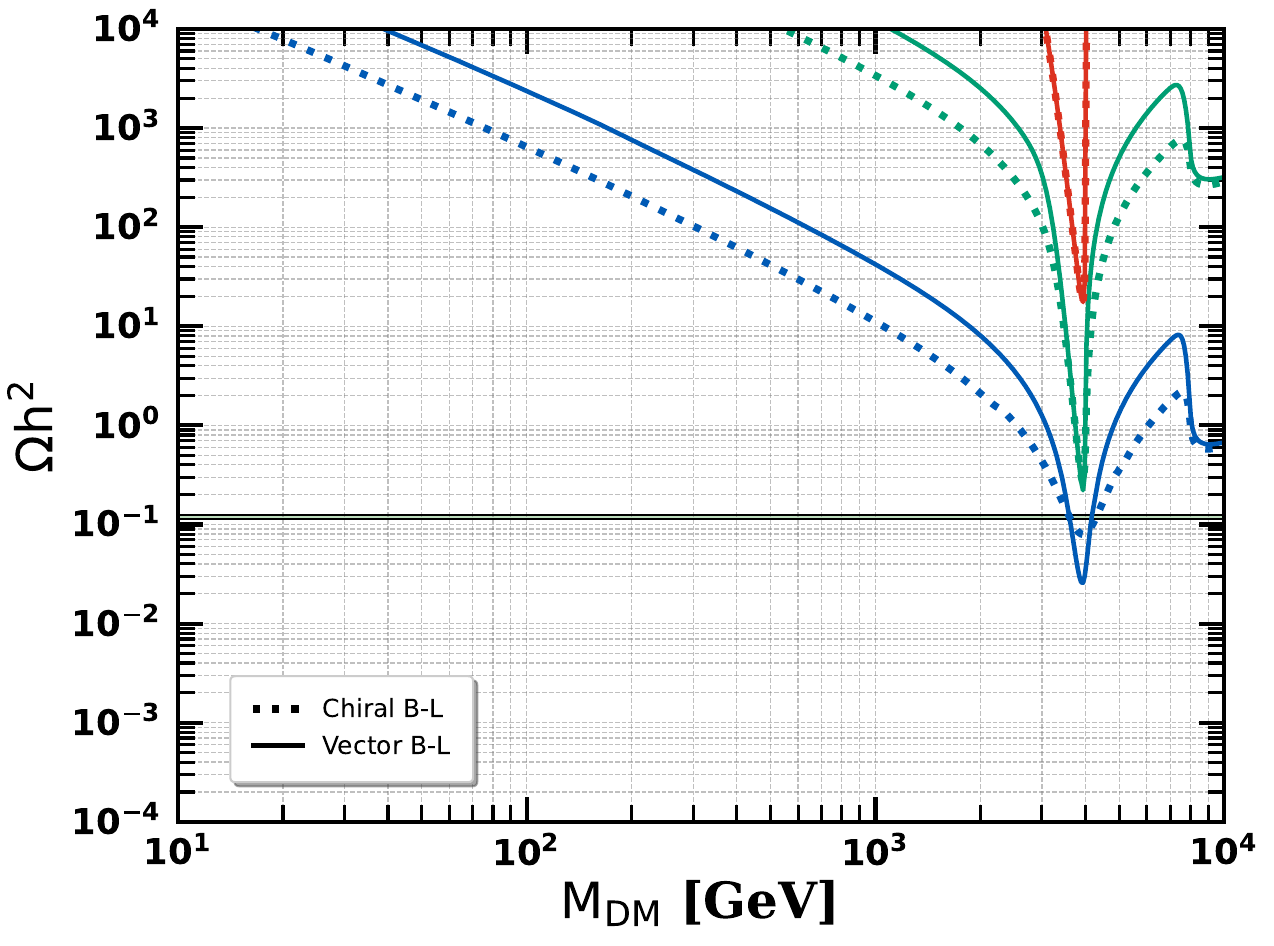}    \includegraphics[width=0.49\linewidth]{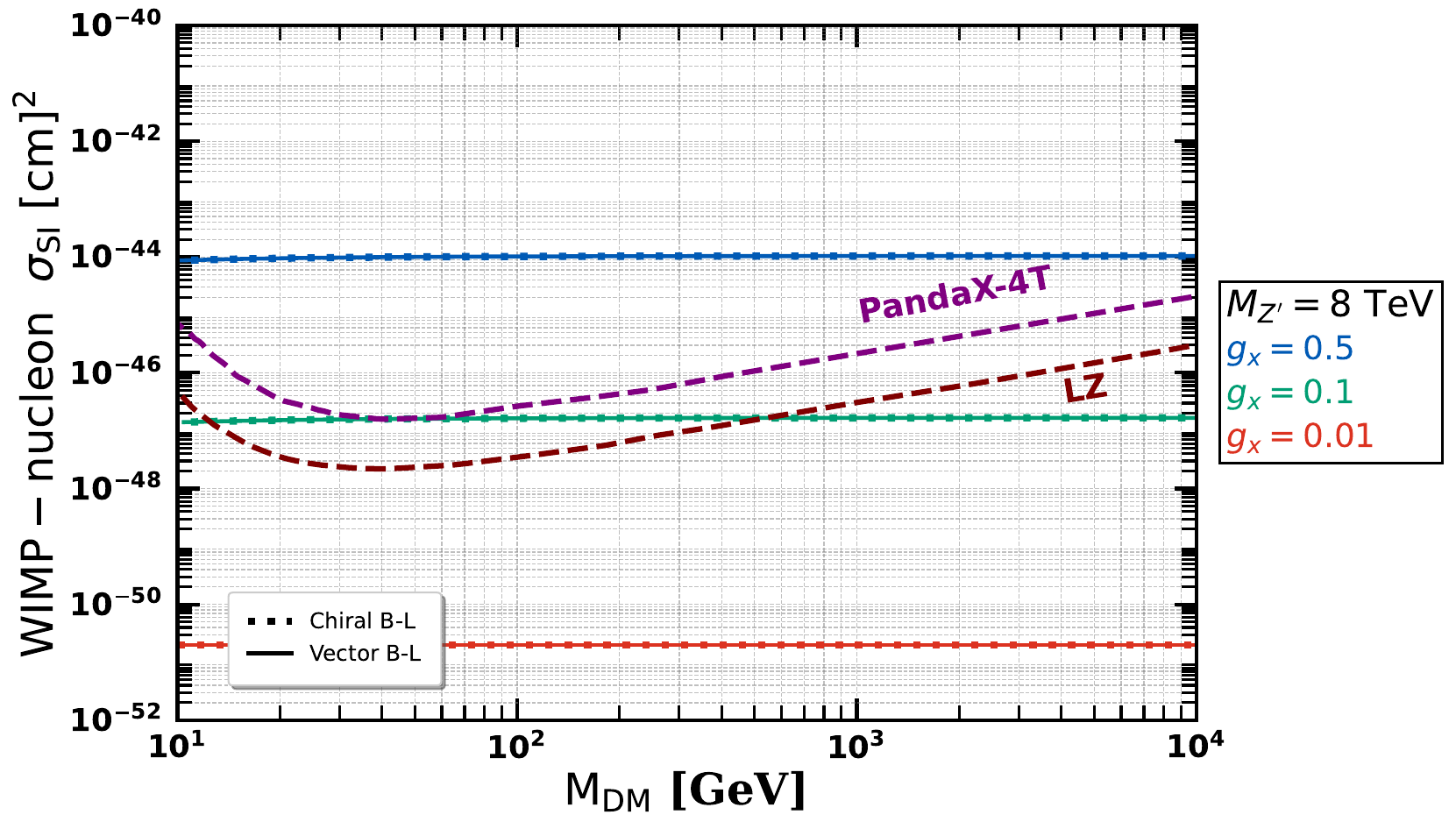}
    \caption{The comparison between chiral and vector $(B-L)$ model is shown. Relic density (left panels) and spin-independent WIMP-nucleon scattering cross section (right panels) as a function of the DM mass $M_{\rm DM}$. The different rows correspond to $Z'$ masses $M_{Z'} = 1,\ 4,\ 8$ TeV (from top to bottom). In each panel, the colored curves represent different values of the gauge coupling $g_x$.}
    \label{fig:OnlyZP2}
\end{figure}

In the main text, we have compared the gauge-mediated DM phenomenology of our model with the chiral $(B-L)$ scenario. For completeness, we also compare the chiral and vector $(B-L)$ realizations in this appendix. Both scenarios involve $Z'$-mediated gauge interactions, while the SM fermions carry the conventional $(B-L)$ charges in both realizations; the two scenarios differ in the $(B-L)$ charge assignments of the right-handed neutrinos. This provides a useful comparison of how the chiral and vector-like charge structures affect the resulting DM phenomenology.
To isolate the effect of these different charge assignments, we consider the limit in which the DM phenomenology is governed exclusively by the $Z'$-mediated gauge interactions. We compare the relic density and spin-independent DM-nucleon scattering cross section for the same representative values of $M_{Z'}$ and $g_{x}$. The corresponding results are shown in Fig.~\ref{fig:OnlyZP2}.
We present the chiral and vector $(B-L)$ models by the dashed and solid curves, respectively. The red, green, and blue curves correspond to $g_x=0.01$, $0.1$, and $0.5$, respectively, while the different rows show the results for $M_{Z'}=1$, $4$, and $8$ TeV from top to bottom. A comparison of the relic density predictions reveals a modest but noticeable difference between the two realizations, particularly in the vicinity of the $Z'$ resonance. This difference originates from the distinct $(B-L)$ charge assignments of the right-handed neutrinos. In the vector $(B-L)$ model, the three right-handed neutrinos carry the universal charges $=(-1,-1,-1)$ whereas in the chiral $(B-L)$ model their charges are $(-4,-4,5)$.
The larger magnitude of the right-handed neutrino charges in the chiral case enhances their coupling to the $Z'$ and consequently increases the total $Z'$ decay width. This broadens the $Z'$ resonance and modifies the resonant annihilation rate, resulting in a comparatively shallower relic density dip than in the vector $(B-L)$ case.

The direct detection predictions exhibit a different behavior. The spin-independent DM-nucleon scattering is mediated by the $Z'$ and depends on the couplings of the $Z'$ to quarks. Since the quark $(B-L)$ charges are identical in the two realizations, the corresponding scattering amplitudes are the same. Consequently, the chiral and vector $(B-L)$ predictions for $\sigma^{\rm SI}$ overlap almost exactly, as seen in the right panels of Fig.~\ref{fig:OnlyZP2}. This comparison therefore cleanly demonstrates that the difference in the relic density behavior originates from the modified $Z'$ decay width and the associated right-handed neutrino charge assignments, whereas the direct detection phenomenology is insensitive to this difference.
Overall, the comparison shows that the chiral $(B-L)$ realization can have an impact on the relic density phenomenology through the invisible and BSM decay modes of the $Z'$, while leaving the tree-level spin-independent direct detection prediction essentially unchanged.


\bibliographystyle{utphys}
\bibliography{references}
\end{document}